\documentclass[prd,twocolumn,nofootinbib,preprintnumbers,floatfix]{revtex4}  % for review and submission
\usepackage[plainpages=false, colorlinks=true, anchorcolor=blue, linkcolor=blue, citecolor=blue, bookmarks=false]{hyperref}
\usepackage{graphicx}  % needed for figures
\usepackage{dcolumn}   % needed for some tables
\usepackage{bm}        % for b}   % for math
\usepackage{natbib}
\usepackage{amsmath}   %%%%   added by me
\newcommand{\rthis}[1]{\textcolor{black}{#1}}
\begin{document}
\newcommand{\apjl}{Astrophys. J. Lett.}
\newcommand{\apjs}{Astrophys. J. Suppl. Ser.}
\newcommand{\aap}{Astron. \& Astrophys.}
\newcommand{\aj}{Astron. J.}
\newcommand{\pasp}{PASP}
\newcommand{\araa}{Ann. Rev. Astron. Astrophys. } %ARA$\&$A}
\newcommand{\aapr}{Astronomy and Astrophysics Review}
\newcommand{\ssr}{Space Science Reviews}
\newcommand{\mnras}{Mon. Not. R. Astron. Soc.}
\newcommand{\apss} {Astrophys. and Space Science}
\newcommand{\jcap}{JCAP}
\newcommand{\pasj}{PASJ}
\newcommand{\pasa}{Pub. Astro. Soc. Aust.}
\newcommand{\physrep}{Physics Reports}

% The following information is for internal review, please remove them for submission

% the following line is for submission, including submission to the arXiv!!
%\hspace{5.2in} \mbox{Fermilab-Pub-04/xxx-E}

\title{Constraints on the  variation of fine structure constant from joint SPT-SZ  and XMM-Newton observations}

\author{Kamal 
\surname{Bora}}
 \altaffiliation{E-mail: ph18resch11003@iith.ac.in}
 
\author{Shantanu \surname{Desai}}%
\altaffiliation{E-mail: shntn05@gmail.com}

%\date{\today}

\begin{abstract}
We search for a variation of the electromagnetic fine structure constant ($\alpha \equiv e^2/\hbar c$)  using a sample of 58 SZ selected clusters in the redshift range ($0.2<z<1.5$) detected by  the South Pole Telescope, along with 
X-ray measurements using the  XMM-Newton observatory.  We use  the ratio of the integrated SZ Compto-ionization to its X-ray counterpart as our observable for this search.
We first obtain a  model-independent constraint on $\alpha$ of about 0.7\%, using the fact that the aforementioned ratio is constant as function of redshift.  
We then look for  logarithmic dependence of  $\alpha$ as a function of  redshift: $\Delta \alpha/\alpha = -\gamma \ln(1+z)$, as this is predicted by  runaway dilaton models. We find that  $\gamma$ = $-0.046 \pm 0.1$, which indicates that  \rthis{there is no logarithmic variation of $\alpha$ as a function of redshift}. We also search for a dipole  variation of the fine structure constant  using the same cluster sample. We do not find any evidence for such a  spatial variation.

\end{abstract}

\affiliation{
 Department of Physics, Indian Institute of Technology, Hyderabad, Kandi, Telangana-502285, India} 
\maketitle

\section{Introduction}
Ever since Dirac's ingenuous argument that Newton's Gravitational Constant could vary with time~\cite{Dirac}, a number of theories beyond General Relativity and the Standard Model of Particle Physics have predicted the  variation of fundamental constants, including the fine-structure constant ($\alpha \equiv e^2/\hbar c$)~\cite{Uzan,Martinisreview}. Therefore, a plethora of searches have been undertaken using both laboratory and astrophysical observations to search for variations of $\alpha$. For theories which predict a variation of $\alpha$, Einstein's equivalence principle is violated in the electromagnetic sector. These theories usually involve  coupling between the electromagnetic part of matter fields and scalar fields~\cite{Bekenstein,Polyakov,Barrow,Barrow2}.

Another strong impetus in searching for  a varying $\alpha$ using different methods, comes from a claimed variation of $\alpha$ using quasar absorption systems~\cite{Webb} with the Keck/HIRES and VLT/UVES telescopes. Subsequently, the same group also argued for a spatial variation in $\alpha$ at 4.2$\sigma$ significance, which  is consistent with a dipole~\cite{Webbspatial,King}. The best-fit position of this dipole is at RA=$(17.5 \pm 0.9)$ hr and Declination=$(-58 \pm 9)^{\circ}$~\cite{Webbspatial,King}. However, other groups  have failed to confirm this result (eg.~\cite{Srianand}). \rthis{Recent studies however indicate that this  dipole variation could be due to the  distortion of the wavelength scale over long wavelength ranges ($\geq 1000$ \AA)~\cite{Whitmore15}. These distortions of the wavelength calibration were demonstrated using asteroid and iodine cell star exposures from both Keck-HIRES and VLT-UVES~\cite{Whitmore15}, and the aforementioned work also showed that simulated quasar spectra can mimic some of the VLT-UVES features seen in ~\cite{King}, although they cannot explain the Keck-HIRES results. Recently, measurements  of zinc and chromium absorption lines in quasars, (which are not sensitive to the aforementioned wavelength distortions) were not consistent with no variations in $\alpha$ within $1\sigma$~\cite{Prochaska16,Prochaska17}. }

 Another motivation in recent years to search for variation in  $\alpha$
comes from the Hubble tension conundrum related to the discrepancy in Hubble constant values between the low-redshift and high redshift probes~\cite{Riess,Bethapudi}. Although current studies indicate that a  variation of $\alpha$ can only play a minor role in resolving the Hubble tension~\cite{Knox,Hart2}, a variation in $\alpha$ could have implications for this problem.

Therefore, a large number of searches for variations of $\alpha$ have   been carried out using a plethora of astrophysical/cosmological probes such as  CMB~\cite{Menegoni,Planck14,Hart,smith19}, Big-Bang Nucleosynthesis~\cite{Ciara}, supermassive black hole at the galactic center~\cite{Ghez},  white dwarf spectra ~\cite{whitedwarf}, strong gravitational lensing~\cite{Cola20}, and also Earth-based measurements using the Oklo natural reactor~\cite{Dyson}, \rthis{atomic clocks~\cite{Godun}} etc. A recent review summarizing the latest results from all  these searches  can be found in ~\citet{Martinisreview} and references therein. All these  other searches  have come up with null results, and failed to corroborate any  claims for a  variation in $\alpha$. However, given the positive claim from one group, it is important to continue searching for a variation of $\alpha$ using multiple  independent sources, as we continue to gather new data.

Here, we use galaxy clusters in testing for variations  in $\alpha$. Galaxy clusters are the most massive, gravitationally collapsed objects in the universe and have proved to be wonderful laboratories for studying cosmology, structure formation, galaxy evolution, neutrino mass, graviton mass, various modified gravity theories etc~\cite{Allen,Vikhlininrev,Desai}.
In the past two decades, a large number of galaxy clusters have been discovered upto very high redshifts thanks to multi-wavelength surveys in optical, X-ray, and Sunyaev-Zeldovich (SZ, hereafter) at mm wavelengths which have mapped out large-area contiguous regions of the universe.
The first test for the  variation of $\alpha$ using cluster SZ observations was implemented by Galli~\cite{galli}, who showed that the ratio of integrated Compto-ionization in SZ ($Y_{SZ})$ to its X-ray counterpart ($Y_X$) scales  as $\alpha^{3.5}$.
A limit  on the variation of $\alpha$ was obtained from the fact this ratio  for  the 2011 Early Planck Release~\cite{planck}) catalog is constant.~\citet{hola1}  then showed that the ratio of gas fraction from SZ and X-ray measurements scales as $\alpha^3$, after assuming that a variation of $\alpha$ leads to a violation of cosmic distance-duality relation (CDDR) ~\cite{Etherington}. They   compared the gas fraction measurements for 29 clusters in the redshift range $0.14<z<0.89$~\cite{Roque}  to constrain $\alpha$ variation at these redshifts. In a follow-up work, Holanda et al~\cite{hola2}  combined the angular diameter distance of galaxy clusters along with luminosity distance measurements from type Ia supernovae  in the redshift range $0.023 < z < 0.784$ to constrain  variations in $\alpha$. A similar idea was thereafter applied to the combination of gas fraction measurements of Atacama Cosmology Telescope selected  clusters  in the redshift range $0.12<z<1.36$ and Type Ia supernovae~\cite{hola3}. 
Martino et al~\cite{martino16D} looked for  spatial evolution of $\alpha$ from Planck SZ data, by looking for variations in the CMB temperature as a function of redshift at the location of the clusters. Colaco et al~\cite{colaco} (C19, hereafter)  carried out a similar analysis as in ~\cite{galli} by looking for variations in the ratio of $Y_{SZ}$ to $Y_X$ as a function of redshift. The main difference with respect to ~\cite{galli}, is   that they assumed (similar to ~\cite{hola1,hola2,hola3}) that a variation in $\alpha$ also leads to a violation of the CDDR.  
Motivated by runaway dilaton models~\cite{damour,damour1,martins},
C19 thereafter modeled the variation in $\alpha$ as a logarithmic function of the redshift, and used
the Planck Early SZ catalog to constrain this variation.
All these searches have failed to find any evidence for the variation of $\alpha$.

In this work, we first implement the same procedure as ~\cite{galli}, to  constrain the model-independent variation of $\alpha$.
Then, similar to  C19, we look for logarithmic variations of $\alpha$ as a function of redshift, by considering  South Pole Telescope-selected  clusters (with joint X-ray and SZ observations) in the redshift range  0.2 $\leq$ z  $\leq$ 1.5. This spans a wider redshift range than in C19, which looked for clusters with $z<0.5$. With the same dataset, we also look for a spatial variation in $\alpha$.

This paper is organized as follows. The basic theory behind the X-ray and SZ observables used for the analysis  is presented in Sect.~\ref{sec:method}.  Our model for the variation of $\alpha$ is described in Sect.~\ref{sec:alphavariation}. The dataset used for our analysis is discussed in Sect.~\ref{sec:data}. A model-independent constraint on variation of $\alpha$ can be found in Sect.~\ref{sec:gallimethod}. Our results for time-varying (using runaway dilaton model) and spatial searches for $\alpha$  can be found in Sect.~\ref{sec:analysis}   and Sect.~\ref{sec:dipolesearch} respectively. We conclude in Sect.~\ref{sec:conclusions}.

\section{Method}
\label{sec:method}
The basic observable which we use for this work is 
the dimensionless ratio of $Y_{SZ}$ to $Y_X$ (after suitable scalings). More precise definitions will be given in forthcoming subsections.
Both $Y_{SZ}$ and $Y_X$ are different proxies for the thermal energy of the cluster and are proportional to cluster mass.
\rthis{They also provide insights on the amount of inhomogeneity and clumping of the intra-cluster medium~\cite{Planelles17}.}
Therefore, studying the relation between the two is important for both astrophysics and cosmology. Consequently, a large number of works~\cite{Bonamente,Andersson10,Bonamente12,More,Rozo,Rozo2,Liu,Chiu18,DeMartino16,Biffi14,Planelles17,Bender,Zhu,Pratt,Henden} have studied the scaling relations between $Y_X$ and $Y_{SZ}$ using both data and simulations to characterize the systematics in the  mass determination as well as any departures from self-similarity. Here, we use this ratio to test for a variation in $\alpha$.
 
 %The scaling relation is given by, 
%\begin{equation}
%    \frac{Y_{SZ} D_A^2}{Y_X C_{XSZ}} =  Constant
%\end{equation}
%which indirectly depends on $\alpha$ and $\eta$. 

\subsection{SZ effect}
The thermal SZ effect refers to  the distortion in the CMB spectrum due to the inverse Compton scattering between the hot gas  present in the intracluster medium and the CMB photons~\cite{sz,birki,carl,SZ18}. Since the SZ effect is a spectral distortion, which is  independent of redshift, it  has become a powerful probe to detect galaxy clusters upto  very high redshifts. In the past decade, there have been  three primary experiments: South Pole Telescope (SPT)~\cite{Carlstrom}, Atacama Cosmology Telescope~\cite{Fowler}, and the Planck satellite ~\cite{Plancksz}, which have carried out a blind SZ survey to detect galaxy clusters upto very high redshifts.  We now discuss    how the SZ signal depends on   $\alpha$, for which we follow the same outline as in C19.

The  distortion measured in SZ experiments is proportional to a parameter, called the Compton parameter $y$, which is given by~\cite{birki,carl}, 
\begin{equation}
    y = \frac{{\sigma_T}k_B}{m_e c^2}  \int n_e T dl
\label{eq:y}    
\end{equation}
Here, $k_B$ is the Boltzmann constant, $c$ is the speed of light, $m_e$ is mass of the electron, $n_e$ is the number density of electrons, $T$ is the electron temperature, and $\sigma_T$ is the Thompson scattering cross section which can be written  in terms of $\alpha$ as,  
\begin{equation}
    \sigma_T = \frac{8\pi}{3} \left(\frac{\epsilon^2}{m_e c^2}\right)^2  = \frac{8\pi}{3}  \left(\frac{\hbar^2\alpha^2}{m_{e}^2 c^2}\right)
\label{eq:sigmat}    
\end{equation}
 Now, the integrated Compton parameter $Y_{SZ}$ (over the solid angle of a galaxy cluster) can be written as,
\begin{equation}
    Y_{SZ} 	\equiv \int_\Omega y {d\Omega} ,
    \label{eq:YSZ}
\end{equation}
where ${d\Omega} = dA/D_A^2$, \rthis{and $D_A$ is the angular diameter distance}. Assuming an ideal gas equation of state given by $P=n_e k_B T$, where $P$ is the pressure of the intracluster medium, one can combine Eq.~\ref{eq:y} and Eq.~\ref{eq:YSZ} to obtain,
\begin{equation}
    Y_{SZ} D_A^2 	\equiv  \frac{\sigma_T}{m_e c^2} \int P dV
\label{eq:yszdasquare}    
\end{equation}
 Since $\sigma_T \propto \alpha^2$ (cf. Eq.~\ref{eq:sigmat}),  we get
\begin{equation}
    Y_{SZ} D_A^2  \propto \alpha^2 .
    \label{eq:yszda}
\end{equation}
If we model the  variation in $\alpha$  as $\alpha(z) \equiv \alpha_{0} \phi(z)$, where $\alpha_{0}$ is the present value of $\alpha$, the fractional variation in $\alpha$  is given by,
\begin{equation}
\frac{\Delta \alpha}{\alpha_0}= \phi(z)-1
\label{eq:deltaalpha}
\end{equation}
Eq.~\ref{eq:yszda} can then be re-written as 
\begin{equation}
    Y_{SZ} D_A^2  \propto \phi(z)^2
    \label{eq:YSZDa2}
\end{equation}

\subsection{X-rays}

At  high temperatures,  gas present in the intracluster medium  emits in X-rays mainly through the thermal
bremsstrahlung process~\cite{Allen}. The thermal energy of the gas can be parameterized by the $Y_X$ parameter~\cite{Kravtsov}, which can obtained from  X-ray surface brightness observations and is given by,
\begin{equation}
    Y_X = M_g(R) T_X
    \label{eq:yxmg}
\end{equation}
where $T_X$ is the X-ray temperature of the gas and $M_g(R)$ is mass of the gas within the radius $R$. Kravtsov et al~\cite{Kravtsov} have shown  that $Y_X$ is strongly correlated with the cluster mass with an intrinsic scatter of about 5-8\%, and hence is a very robust proxy for the cluster mass.

%\begin{equation}
%    M_g(R) = \mu_e m_p \int n_e dV
%\end{equation}
%Here $m_p$ is the proton mass and $\mu_e$ represents the mean molecular weight of the electrons. 
The gas mass $M_g(R)$ is scales with  $\alpha$ as ~\cite{Holanda11,galli,colaco},
\begin{equation}
      M_g (<\theta) \propto \phi(z)^{-3/2} D_L D_A^{3/2},
      \label{eq:mg2}
\end{equation}
where $D_L$ is the luminosity distance. From Eq.~\ref{eq:mg2}, we can  see that $M_g(R)$
depends upon both $D_L$ and $D_A$. Both of them are
linked by CDDR, $D_L = (1+z)^2 D_A$~\cite{Etherington}. As shown in ~\cite{hees,rodrigo19}, any variation in $\alpha$  is intertwined with the  violation of CDDR, which must be included in  any searches for variation of $\alpha$~\cite{colaco}. Similar to C19, if we parameterize  a violation of CDDR  using $D_L$ = $\eta(z)$ $(1+z)^2$ $D_A$, then  the variation of  $Y_X$ scales according to,
\begin{equation}
        Y_X   \propto \phi(z)^{-3/2} \eta(z)
        \label{eq:Yx}
\end{equation}
As argued in C19 (and references therein) for modified theories of gravity where the scalar field couples to the electromagnetic  sector and breaks the equivalence principle, $\alpha(z)$ is related to $\eta(z)$ according to  $\alpha(z)=\eta(z)^2$~\cite{Cola20,hees}. Therefore,  Eq.~\ref{eq:Yx} can be recast as
\begin{equation}
Y_X   \propto \phi(z)^{-1}
\label{eq:Yx2}
\end{equation}

\section{Model for $\alpha$ variation}
\label{sec:alphavariation}
\iffalse
From Eq.~\ref{eq:YSZDa2} and Eq.~\ref{eq:yxmg}, we get 

\begin{equation}
    \frac {\Delta\alpha}{\alpha} (z) \equiv \frac {\alpha(z) -\alpha_0}{\alpha_0} =  \phi(z) - 1 =  \eta(z)^2 - 1
\end{equation}

Now, dependency of Eq.~\ref{eq:YSZDa2} and Eq.~\ref{eq:Yx} on $\phi(z)$ as,

\begin{equation}
    Y_{SZ} D_A^2  \propto \phi^2
\end{equation}

\begin{equation}
    Y_X \propto \phi^{-1}
\end{equation}

~\cite{galli} found the different result which is $Y_{SZ} D_A^2  \propto \phi(z)^{-3/2}$ because they have not accounted the dependency of $Y_X$ on CDDR.
\fi

The dimensionless ratio of $Y_{SZ} D_A^2$  to $Y_X$   can be combined  from Eq.~\ref{eq:yszdasquare} and Eq.~\ref{eq:Yx}~\cite{galli}
\begin{equation}
    \frac{Y_{SZ} D_A^2}{Y_X} = C_{XSZ} \frac{\int n_e(r) T(r) dV }{T(R) \int n_e(r) dV},
    \label{eq:ratio}
\end{equation}
where 
\begin{multline}
     C_{XSZ} = \left(\frac{\sigma_T}{m_e m_p \mu_e c^2}\right) \\ \approx 1.416 \times 10^{-19} \left( \frac{ Mpc^2}{M_\odot keV}\right)
\end{multline}
%\end{equation}

The numerator and denominator in Eq.~\ref{eq:ratio} are different proxies for the thermal energy of the cluster.   Simulations  show that the ratio in Eq.~\ref{eq:ratio} is expected to be a constant with redshift with a scatter of approximately 5-15\%~\cite{stanek,fabjan,kay,Biffi14,Planelles17}. For clusters with isothermal or universal temperature profile~\cite{Norman}, this  ratio should be  constant with redshift and  equal to unity~\cite{galli,colaco}. Therefore, assuming no new Physics, there should not be any variation for this ratio as a function of  redshift.

Following C19, we combine Eq.~\ref{eq:YSZDa2} and Eq.~\ref{eq:Yx2} to rewrite the ratio in Eq.~\ref{eq:ratio} in terms of the variation in  $\alpha$  as,
\begin{equation}
    \frac{Y_{SZ} D_A^2}{Y_X C_{XSZ}} =  C \phi^3
\label{eq:ratiowithphi}
\end{equation}
where $C$ is an unknown constant, which encapsulates all the cluster astrophysics in  the ratio in 
Eq.~\ref{eq:ratio}. A value of $C$ close to one further indicates that the cluster gas has an isothermal profile. Similar to C19 (see also ~\cite{hola1,hola2,hola3}), we use runaway dilaton models~\cite{damour,damour1,martins} to parameterize the variation in $\phi$ as
\begin{equation}
   \phi(z) =  1 -\gamma  \ln(1+z)
   \label{dilatonphi}
\end{equation}
In the dilaton model proposed by Damour et al~\cite{damour,damour1}, Eq.~\ref{dilatonphi} was derived  by assuming that the velocity of the dilaton field is constant in both matter and dark-energy  eras (with different constants for both). With these assumptions, Eq.~\ref{dilatonphi}
is valid only in the dark energy dominated era and needs to be augmented by an additional constant term in the matter dominated era. However, De Martinis et al~\cite{martins,Martinisreview} have shown that Eq.~\ref{dilatonphi} can also be obtained by directly integrating the Friedman and scalar field equations for the runaway dilaton model, followed by linearizing the field evolution. As pointed out in ~\cite{martins}, this approximation is valid for low redshifts, but breaks down at large  redshifts (close to $z=5$). At these redshifts, the variation with respect to $z$ is model-dependent, depending on the couplings of dilaton field to baryons, dark matter, and dark energy. These variations have been calculated for the different couplings, and can be found in ~\cite{erminia} and can be parametrized by polynomial functions of $z$.

Since our sample extends only to redshift of 1.5, we use Eq.~\ref{dilatonphi} to constrain any putative variation in $\alpha$, as this function has also been used in previous works. However, it is straightforward to extend this analysis to any other parametric model for the variation of $\alpha$.
From Eq.~\ref{dilatonphi}, the fractional variation in $\alpha$ can be written as 
\begin{equation}
   \frac{\Delta \alpha}{\alpha} =  -\gamma  \ln(1+z)
\label{eq:deltaalphagamma}   
\end{equation}
Therefore,  the relation in Eq.~\ref{eq:ratiowithphi} now becomes
\begin{equation}
    \frac{Y_{SZ} D_A^2}{Y_X C_{XSZ}} =  C(1 - \gamma  \ln(1+z))^3
    \label{eq:compare}
\end{equation} 

We now fit for this relation using  the SPT data. \rthis{We note that we have used $D_A$  (needed for evaluating Eq~\ref{eq:compare}) using the Planck 2018 $\Lambda CDM$ cosmological parameters~\cite{planck18} and evaluated using the {\tt astropy}~\cite{astropy} module. However, that the Friedman equations themselves get corrections due to the  $H^2{\phi'}^2$ term in the presence of a dilaton field~\cite{damour,damour1,Martinisreview}, where the prime indicates the derivative with respect to  the logarithm of the scale factor. As shown in ~\citet{Martinisreview}, $\frac{1}{H}  \frac{\dot{\alpha}}{\alpha} \sim \frac{\alpha_{had}}{40}\phi'$,  where $\alpha_{had}$ is the coupling of the dilaton field to hadronic matter.
Based on reported limits of $\frac{\dot{\alpha}}{\alpha}<10^{-17}$/yr from atomic
clocks~\cite{Godun}, one expects  negligible corrections to Friedmann equations from the dilaton couplings, over the redshift range used for the SPT sample. Therefore, it is reasonable to neglect the contribution of the dilaton field to Friedmann equation, and to use the standard Friedmann equation to compute angular diameter distance.}

\begin{figure}[h]
    %\centering
    \includegraphics[width=0.5\textwidth]{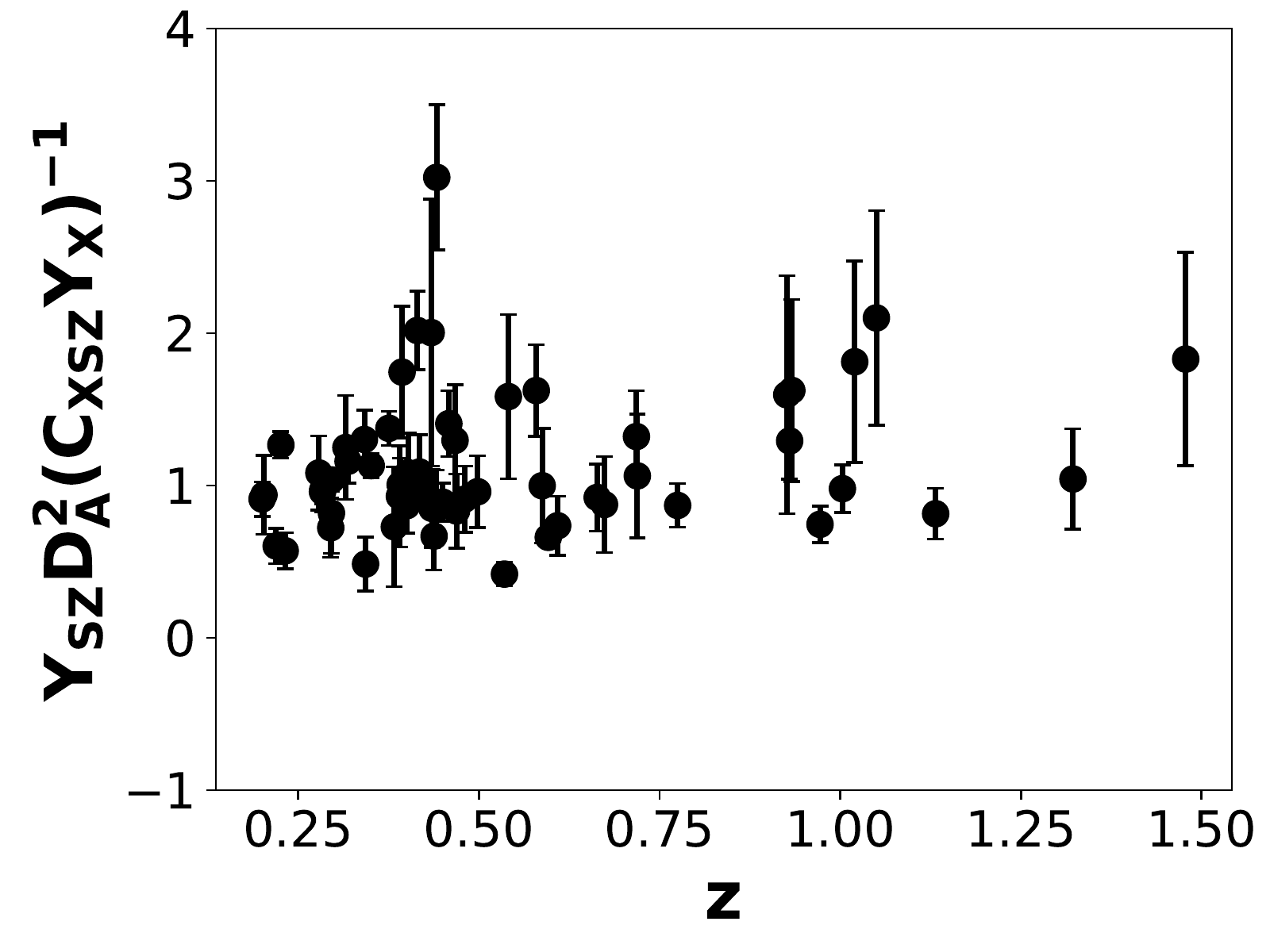}
    \caption{The observed values of  $Y_{SZ}$ $D_A^2$ / $C_{XYZ}$ $Y_X$  for the SPT galaxy cluster sample together  with X-ray measurements from XMM-Newton observations~\cite{Bulbul}.}
    \label{fig:f1}
\end{figure}

\section{SPT Cluster Sample}
\label{sec:data}
For this work, we use joint X-ray  and SZ data for 58 SPT galaxy clusters~\cite{Bulbul}.
The SZ data have been obtained by  SPT, which is a 10~m telescope at the South Pole, that  has imaged the sky at three different frequencies, viz. 95 GHz, 150 GHz and 220 GHz~\cite{Carlstrom}. SPT carried out a 2500 square degree survey between 2007 and 2011 to detect galaxy clusters using the SZ effect.
This SPT-SZ survey detected 516 galaxy clusters with mass threshold of $3 \times 10^{14} M_{\odot}$ upto redshift of  1.8~\cite{Bleem15}. Detailed properties of the SPT  clusters (confirmed until 2015)  are discussed in~\cite{Bleem15}. Their redshifts  have been obtained using a dedicated optical follow-up campaign, consisting of  pointed spectroscopic and photometric observations~\cite{Song,Ruel,Bayliss}, and also using data from surveys mapping out contiguous regions of the sky such as BCS~\cite{Desai12} and DES~\cite{Saro}.  
For every SPT cluster,  $Y_{SZ}$ has been estimated by averaging in a cylindrical volume within an aperture  radius of 0.75$'$ ~\cite{Bleem15}.  These cylindrically averaged $Y_{SZ}$ values for the  confirmed SPT clusters are reported in~\cite{Bleem15} and recently updated in ~\cite{Bocquet19}. 
In order to search for a variation of $\alpha$, one needs to compare the $Y_X$ measured within a radius $R$, with the  $Y_{SZ}$ obtained by averaging over a spherical volume within the same radius $R$~\cite{galli,hola1}. Since the SPT  $Y_{SZ}$ are obtained from cylindrical averaging over an angular aperture of 0.75$'$, we first need to convert these values to spherically averaged $Y_{SZ}$~\cite{Arnaud} within the same radius at which $Y_X$ was measured. 
To do this conversion we follow the prescription in~\cite{Arnaud} (see also ~\cite{Melin}). The cylindrical and spherically averaged $Y$ can be related using:
\begin{widetext}
\begin{eqnarray}
Y_{cyl}(R_1) &=& Y_{sph}(R_b) - \frac{\sigma_T}{m_e c^2} \int_{R_1}^{R_b} 4\pi P(r) \sqrt{r^2-R_1^2}r dr 
\label{eq:cylY} \\
Y_{sph}(R_2) &=& \frac{\sigma_T}{m_e c^2} \int_{0}^{R_2} 4\pi P(r) r^2 dr
\label{eq:sphY}
\end{eqnarray}
\end{widetext}

\noindent where $Y_{cyl}$ is the SZ signal within a cylindrical aperture of radius $R_1$; $R_b$ is the radial extent of the cluster; and $P(r)$ is the gas pressure in the intra-cluster medium. Similarly, $Y_{sph}$ is the corresponding integrated SZ flux within a sphere of radius $R_2$. To do the conversion, we use the Universal Pressure Profile to model  $P(r)$~\cite{Arnaud}. We choose $R_b=10 R_{500}$. Since the SPT $Y_X$ are reported at $R_{500}$ and $Y_{SZ}$ values calculated for an aperture of 0.75$'$, we assume $R_1=0.75'D_A$ in Eq.~\ref{eq:cylY} and $R_2 = R_{500}$ in Eq.~\ref{eq:sphY}. Therefore, from Eq.~\ref{eq:cylY} and Eq.~\ref{eq:sphY},  we can estimate the ratio of $Y_{sph}(R_{500})$ to $Y_{cyl}(0.75' D_A)$, which is then used to estimate $Y_{sph}(R_{500})$. As pointed out by C19,  $Y_{SZ}$ values could also be affected by  a modification of the adiabatic evolution of the CMB temperature as a function of redshift. The SPT collaboration has  looked for such a violation, and their results are consistent with the standard model of CMB temperature variation with  redshift~\cite{Saro13}.

SPT has also been undergoing a massive X-ray followup campaign using both the Chandra and XMM-Newton telescopes. Some details of the Chandra-based followups can be found in~\cite{McDonald13,McDonald14}. Previous studies for the scaling relation between SPT $Y_{SZ}$  and $Y_X$ (with Chandra measurements) as well as other Physics results have been reported in ~\cite{Andersson10,Semler,McDonald13,McDonald14,McDonald17,Chiu18,McDonald19}.

Here, we shall use XMM-Newton followup observations of 58 SPT clusters described in Bulbul et al~\cite{Bulbul}.
These XMM-Newton observations have been obtained using a combination of targeted X-ray followup programs, led  by SPT collaboration members and also other non-SPT based small programs.  The redshift range  of this sample is given by  0.2 $<$ $z$ $<$ 1.5. The exposure time for each cluster is $\mathcal{O} (10)$ ks. More details of the observations and XMM-Newton data reduction can be found in~\cite{Bulbul}. $Y_X$ values for all the 58 clusters have been provided at $R_{500}$ in ~\cite{Bulbul}. \rthis{The temperatures used for SPT $Y_X$ evaluations are emission-weighted temperatures~\cite{Bulbul}}. Note that these ~\cite{Bulbul} provide  both the core-excised as well as the core-included $Y_X$ data for all the 58 clusters. For  our analysis, we combined these  \rthis{core-included} $Y_X$ estimates along with the spherically averaged $Y_{SZ}$ (converted from those in~\cite{Bleem15}), to construct the  ratio  $(Y_{SZ}$ $D_A^2$/ $C_{XYZ}$ $Y_X)$ as a function of redshift. This variation  is shown in Fig.~\ref{fig:f1}. By eye, we see no trends with redshift, thus indicate that the ratio is constant as a function of redshift. We quantify this in the next sections.

\section{Model-independent constraint}
\label{sec:gallimethod}
We first report a model-independent constraint on the variation of  $\alpha$, along the same lines as Galli~\cite{galli}, by using the fact that the ratio of SZ to X-Ray Compto-ionization ratio is constant as a function of redshift.

Similar to ~\cite{galli}, we use the modified weighted least-squares method (MWLS) to calculate the new weighted mean for the ratio and its intrinsic scatter.
From the ordinary weighted least-squares method we find from  the measurements in Fig.~\ref{fig:f1} that $\ln (\frac{Y_{SZ} D_A^2}{Y_X C_{XSZ}})=0.05 \pm 0.02$, with $\chi^2/dof= 237/57$.
Therefore, using the MWLS, we keep both the weighted mean and the intrinsic scatter as  free parameters.  For each value of the new weighted mean, the intrinsic scatter is chosen so that the reduced $\chi^2$ is equal to one. Among these values, the weighted mean  which gives the minimum value for the intrinsic scatter was chosen as the new weighted mean. Using this procedure, we get $\ln (\frac{Y_{SZ} D_A^2}{Y_X C_{XSZ}})=0.046 \pm 0.02$, with an intrinsic scatter of 29\%. This corresponds to $(\frac{Y_{SZ} D_A^2}{Y_X C_{XSZ}})=1.05 \pm 0.02$

The uncertainty in the variation of $\alpha$ is then given by~\cite{galli} 
\begin{equation}
\frac{\sigma(\alpha)}{\alpha_0} = \frac{1}{X} \frac{\sigma (Y_{SZ} D_A^2/Y_X C_{XSZ})}{(Y_{SZ} D_A^2/Y_X C_{XSZ})}    
\label{gallieq}
\end{equation}
where $X$ is equal to 3  or 3.5, depending on whether the   theoretical model for $\alpha$ variation violates CDDR or not. From this we get $\frac{\sigma(\alpha)}{\alpha_0}$ $\sim$ 0.7\% (violation of CDDR) or 0.6\% (no violation of CDDR). This is about the same level of precision as that obtained using the Planck ESZ sample~\cite{galli}.
Similar to ~\cite{galli}, this limit also assumes no astrophysical evolution of the X-ray to SZ Compto-ionization ratio. However, here we have included the error in cosmological parameters while calculating the observed Compto-ionization ratio, unlike ~\cite{galli}.

\section{Constraints on runaway dilaton models}
\label{sec:analysis}
To look for variations in $\alpha$, we fit  the observed ratio (Eq.~\ref{eq:ratiowithphi}) for the SPT cluster sample to the model in Eq.~\ref{eq:compare}, and obtain the best-fit values of $C$ and $\gamma$ by maximization of the log-likelihood.
The log-likelihood function ($\mathcal{L}$) used to test for the variation in $\alpha$ can be written as,

\begin{widetext}
\begin{equation}
  -2 \ln \mathcal{L}  = \sum_{i} \ln 2\pi{\sigma_{i}^2} +  \sum_{i=1}^{N}\frac{(\phi_{obs,i} - C (1 -\gamma  \ln(1+z))^3 )^2}{\sigma_{i}^2}
 \label{eq:likelihood} 
\end{equation} 
\end{widetext}
%where,
%\begin{equation}
%   \chi^2 = \sum_{i=1}^{n}\frac{(\phi_{obs,i} - C (1 -\gamma  \ln(1+z))^3 )^2}{\sigma_{i}^2}
%\end{equation}                                         
where $\phi_{obs,i}$ denote the observed values of ratio in Eq.~\ref{eq:ratio} for SPT clusters, $N$ is the total number of clusters, and $\sigma_i$ denotes the total error calculated as 
\begin{equation}
     \sigma_{i}^2 =  \sigma_{\phi}^2  +  \sigma_{int}^2
     \label{eq:sigma}
\end{equation}
Here,  $\sigma_{\phi}$, which is the error in  $\phi_{obs,i}$ is  obtained by propagating the error in $Y_X$,  $Y_{SZ}$, and $D_A$. We also added an intrinsic scatter term ($\sigma_{int}$) in quadrature, which we keep as a free parameter, 
while  maximizing the log-likelihood.  \rthis{We note that the intrinsic scatter is often added as a free parameter in linear regression problems, since the relation between $y$ and $x$
may not always be  100\% deterministic even for noise-free observations, for a variety of reasons~\cite{Kelly,Hogg10}. The use of intrinsic scatter is  ubiquitous  in galaxy cluster astrophysics to quantify  how well a  given scaling relation is obeyed (eg. ~\cite{Pratt,Liu,Chiu20}). A small value for this quantity indicates negligible deviations from the scaling relation
and vice-versa. Adding the intrinsic scatter term as a free parameter  to the observational errors is similar to the method adopted in ~\cite{Tian}, who used a similar expression to test the tightness of the radial acceleration relation for galaxy clusters. We note however that our treatment of systematic error  differs from C19, who used a fixed intrinsic scatter of 17\%.}
For calculating $D_A$ and its error, as mentioned earlier,  we used the Planck 2018 cosmological parameters~\cite{planck18} ($H_0$ = $67.4\pm0.5$ km/sec/Mpc and $\Omega_m$ = $0.315\pm 0.007$).

\begin{figure*}
    \centering
    \includegraphics[width=20cm, height=17cm]{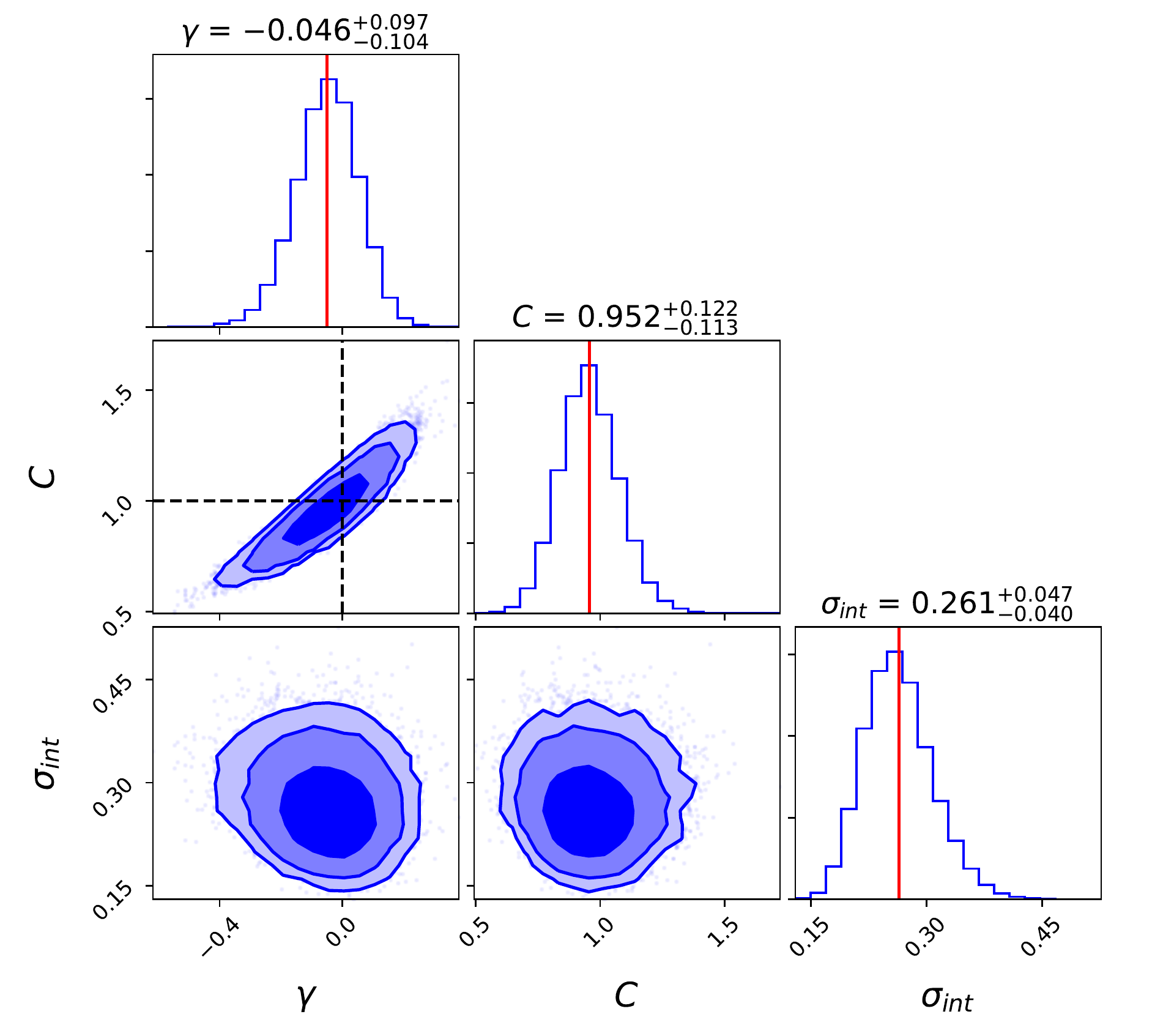}
    \caption{Constraints on $\gamma$ , $C$ (defined in Eq.~\ref{eq:ratiowithphi}) and $\sigma_{int}$ (defined in Eq.~\eqref{eq:sigma}), which are  obtained by maximizing the likelihood in Eq.~\ref{eq:likelihood} using the {\tt emcee} module. The plots along the diagonal are the 1-D marginalized likelihood distributions. The red line represents the mean value of the distribution quoted right above the histogram. The off-diagonal plots are the two-dimensional marginalized constraints showing the 68\%, 95\%, and 99\% credible regions, obtained using the {\tt Corner} module. Our results are consistent with no variation of $\alpha$ and an isothermal profile for clusters.}
    \label{fig:f2}
\end{figure*}

We maximize the likelihood using the {\tt emcee} MCMC sampler~\cite{emcee}.
The 68\%, 95\%, and 99\% confidence level plots along with the marginalized one-dimensional  likelihoods for each of the three parameters  are displayed in Fig.~\ref{fig:f2}. The best fits which we get are  $\gamma$ = $-0.046_{-0.1}^{+0.097}$ and $C$ = $0.95_{-0.11}^{+0.12}$, with an intrinsic scatter of about 26\%. Therefore, our results indicate that there is  no evolution of $\gamma$ with redshift, implying no variation in  $\alpha$.  Furthermore, since the best-fit value of $C$ is consistent with 1.0 within 1$\sigma$ (similar to C19), we conclude that our   cluster sample can be adequately described by the isothermal temperature profile.

\rthis{We note that simulations predict an intrinsic scatter of 5-15\%~\cite{stanek,fabjan,kay,Biffi14,Planelles17}. Our estimated intrinsic scatter of 26\% is larger than these estimates. Other observational results for the intrinsic scatter in $Y_X-Y_{SZ}$ relation range from negligible scatter of 5\%~\cite{More}, to scatters  between 15-20\%~\cite{planck,Zhu} which are comparable to our estimate,  and also  extremely large scatter of 60-90\%~\cite{DeMartino16,Pratt}. These estimates also depend upon  the sample used (including significant differences between cool-core and non cool core~\cite{Zhu}),  as well as  the regression method employed. We discuss some possible reasons for our large scatter compared to the expectations from simulations.}

\rthis{~\citet{Biffi14} showed using   the MUSIC dataset of simulated clusters that  the emission-weighted temperatures show deviations of about 20\%, compared to spectroscopic-like  or mass-weighted temperatures, which show deviations of only about 10\% and 5\% respectively. The aforementioned work also showed that the $Y_X-Y_{SZ}$ scaling relation shows a scatter of 5\% for $Y_X$ computed
using mass-weighted temperatures, which is comparable to the deviations with respect to  the true temperatures. Therefore, one would expect a scatter of about 20\%, if $Y_X$
was estimated using emission-weighted temperatures.
Since the $Y_X$ estimates were obtained using the emission-weighted temperatures, that could be one possible reason for our large scatter. Other possibilities could be because of the use of Universal Pressure Profile~\cite{Arnaud}, which is known to overestimate the thermal SZ signal by upto 15-20\% depending upon the angular aperture~\cite{DeMartino16}. However, a detailed investigation of these and other causes for the large intrinsic will be deferred to future works. }

Table.~\ref{tab:example_table} shows the constraints on $\gamma$ from different cosmological  observations along with our results. We note that the most stringent bound on $\gamma$ are however obtained from tests of violations of weak equivalence principle and are $\mathcal{O}(10^{-6})$~\cite{Vacher}.

% Example table
\begin{table*}
	\centering
	\begin{tabular}{lccccr} % six columns, alignment for each
		\hline
		\textbf{Data Set}  & \boldmath{$\gamma$}  & \textbf{Reference} \\
		\hline
		Only Gas Mass Fractions & $+0.065\pm0.095$ & ~\cite{hola1} \\
	    Angular Diameter Distance + SNe Ia & $-0.037\pm0.157$ & ~\cite{hola2} \\
	   % Gas Mass Fractions + SNe Ia & $+0.008\pm0.035$ & ~\cite{hola3} \\
	    %Gas Mass Fractions + SNe Ia & $+0.018\pm0.032$ & ~\cite{hola3} \\
	    Gas Mass Fractions + SNe Ia & $+0.010\pm0.030$ & ~\cite{hola3} \\
	    %Gas Mass Fractions + SNe Ia & $+0.030\pm0.033$ & ~\cite{hola3} \\
	    ${Y_{SZ} D_A^2}/{C_{XYZ}Y_X}$ & $-0.15\pm0.10$ & ~\cite{colaco} \\
	    Strong Gravitational Lensing + SNe Ia & $-0.013_{-0.09}^{+0.08}$  & ~\cite{Cola20} \\
	   
	    $\mathbf{{Y_{SZ} D_A^2}/{C_{XYZ}Y_X}} $ & $\mathbf{-0.046_{-0.104}^{+0.097}}$ &   \textbf{This work} \\
		\hline
		
	\end{tabular}
		\caption{\label{tab:example_table} A summary of the current astrophysical/cosmological constraints from different works (including the result from this work) on a possible time evolution of $\alpha$ for a class of dilaton runaway models ($\phi$ = 1-$\gamma$ $\ln$(1 + $z$)). 
}

\end{table*}

\section{Search for Dipole variation of $\alpha$} \label{sec:dipolesearch} 

 To confirm the dipole variation of $\alpha$ found using quasar data~\cite{Webbspatial}, Galli~\cite{galli} also searched for a spatial variation of $\alpha$. Due to  their different redshift and spatial distributions  compared to quasars, clusters offer a complementary probe to test the claim in ~\cite{Webbspatial}. The model for the  dipole variation posited in~\cite{galli} is given by 
\begin{equation}
   \frac {\Delta\alpha}{\alpha} =  A r cos(\theta)
\end{equation}
where $A$ represents the dipole amplitude, $\theta$ is the angular separation between clusters and best-fit dipole position found in ~\cite{Webbspatial}, $r$ is the lookback time which is given by,
\begin{equation}
    r =  \int_0^z \frac{c dz^{'}} { H(z^{'})} 
\end{equation}
For a flat $\Lambda$CDM universe, $H(z^{'})$ is given by: 
\begin{equation}
    H(z^{'}) =  H_0 \sqrt{\Omega_m (1+z^{'})^3 + \Omega_{\Lambda}}
\end{equation}

\begin{figure}
    %\centering
    \includegraphics[width=0.5\textwidth]{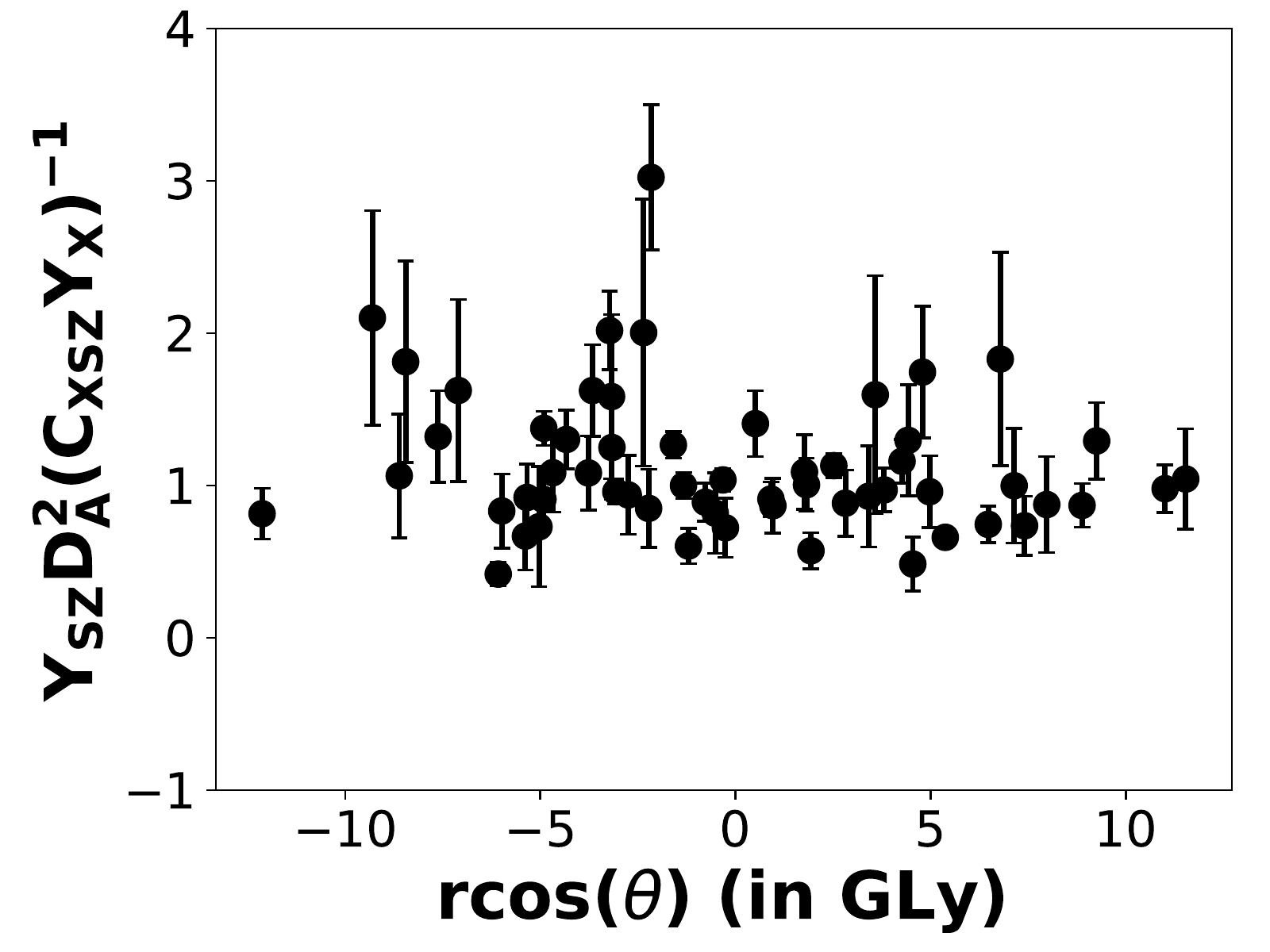}
    \caption{$Y_{SZ}$ $D_A^2$ / $C_{XYZ}$ $Y_X$ as a function of $r cos(\theta)$ for the SPT galaxy cluster sample using XMM-Newton observations~\cite{Bulbul}.}
    \label{fig:f3}
\end{figure}

Therefore, Eq.~\ref{eq:ratio} becomes
\begin{equation}
    \left(\frac{Y_{SZ} D_A^2}{Y_X C_{XSZ}}\right)^{1/3} =  a+b rcos(\theta), 
\label{eq:dipoleeq}
\end{equation}
where $a$ = $Y_{0}^{1/3}$ and $b$ = $A$ $Y_{0}^{1/3}$;   $Y_0$ refers to a  fixed reference value 
for $\left(\frac{Y_{SZ} D_A^2}{Y_X C_{XSZ}}\right)$;  and $\cos(\theta)$ is the angular position between every SPT cluster  and the best-fit dipole position as reported in ~\cite{Webbspatial}.  Note that we have $1/3$ in the exponent of Eq.~\ref{eq:dipoleeq}, unlike $1/3.5$
in~\cite{galli}, as we have assumed (similar to the analysis in Sect.~\ref{sec:analysis}) that a variation in $\alpha$ also leads to a violation of CDDR. Fig.~\ref{fig:f3} shows the ratio $\frac{Y_{SZ} D_A^2}{C_{XYZ} Y_X}$ as a function of $r cos(\theta)$ for the SPT cluster sample.

\begin{figure*}
    \centering
    \includegraphics[width=20cm,height=17cm]{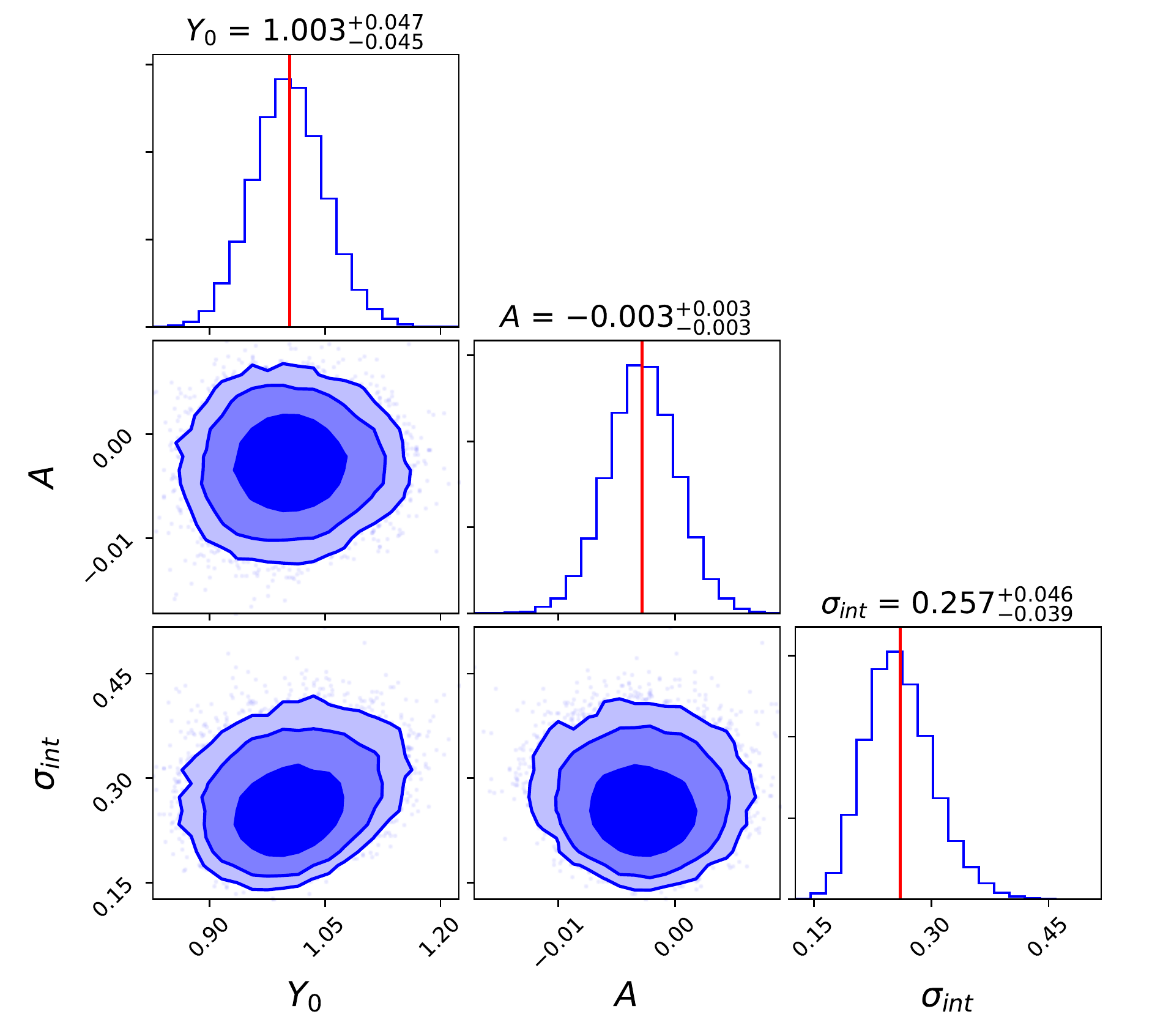}
    \caption{Constraints on $Y_0$, $A$ and $\sigma_{int}$ (defined in Eq.~\ref{eq:dipoleeq}) used to model a spatial variation in $\alpha$. Eq.~\ref{eq:dipoleeq} assumes that CDDR is also violated when $\alpha$  varies. The plots along the diagonal are the one-dimensional marginalized likelihood distributions. The off-diagonal plots are the two-dimensional marginalized constraints showing the 68\%, 95\%, and 99\% credible regions.}
    \label{fig:f4}
\end{figure*}

To get the best-fit value of $Y_0$ and $A$, we maximize a log-likelihood similar to that in Eq.~\ref{eq:likelihood}, after  including an intrinsic scatter. Again, we have used the  {\tt emcee} MCMC sampler  for the optimization. From this analysis, we obtained $Y_0$ = $1.0_{-0.045}^{+0.047}$ and $A$ = $-0.003_{-0.003}^{+0.003}$ $(GLy^{-1})$ with an intrinsic  scatter of about 26\%. The likelihood distributions of our parameters (after assuming violation of CDDR)  along with 2-D marginalized  contours are displayed in Fig.~\ref{fig:f4}.
To compare our results with ~\cite{galli}, We also repeated our analysis for dipole variation of $\alpha$, assuming no violation in CDDR. So we redid the analysis by replacing the exponent of 3 in Eq.~\ref{eq:dipoleeq} with 3.5. With this assumption, we find $Y_0$ = $1.0_{-0.046}^{+0.048}$ and $A$ = $-0.002_{-0.003}^{+0.002}$ $(GLy^{-1})$ with an intrinsic  scatter of about 26\%. Therefore,  our results are also consistent with no dipole variation.  A comparison of parameter $A$ with the  previous  studies in literature is summarized in Table.~\ref{tab:example_table_var}.

% Example table
\begin{table*}
	\centering
	\begin{tabular}{lccccr} % six columns, alignment for each
		\hline
		\textbf{Data Set}  & \boldmath{$A(GLy^{-1})$} & \textbf{Reference} \\
		\hline
		Quasar & ($1.1\pm0.25$) $\times$ $10^{-6}$ & ~\cite{Webbspatial} \\
	    Planck ESZ Clusters & ($-5.5\pm7.9$ )$\times$ $10^{-3}$ & ~\cite{galli} \\
	    \textbf{SPT Clusters} & ($\mathbf{-3.0\pm  3.0 )\times 10^{-3}}$&  \textbf{This work} \\
	    \textbf{SPT Clusters (without CDDR violation)} & ($\mathbf{-2.0_{-3.0}^{+2.0})\times 10^{-3}}$&  \textbf{This work} \\

		\hline
		
	\end{tabular}
\caption{\label{tab:example_table_var}. A comparison of different studies showing the dipole variation of $\alpha$. We note that the study in ~\cite{galli} using Planck ESZ clusters does not assume a  violation of CDDR.
}

\end{table*}

\section{Conclusions}
\label{sec:conclusions}
We search for a variation in $\alpha$ as a function of redshift, using the dimensionless ratio of the integrated Compto-ionization to its X-ray counterpart ${Y_{SZ} D_A^2}/{C_{XYZ}Y_X}$. For this study, we use  data from  58 SPT-SZ selected  galaxy clusters~\cite{Bleem15} in the redshift range 0.2 $\leq$ $z$ $\leq$ 1.5, along with X-ray measurements from XMM-Newton~\cite{Bulbul}. We first do a model-independent test for the variation of $\alpha$ (similar to ~\cite{galli}), by using the  fact that this ratio is constant as a function of redshift. The variation of $\alpha$ is constrained to within 0.6-0.7\% in the redshift range  ($0.2<z<1.5$).

Then, similar to C19, we assumed that a variation of $\alpha$ also leads to a violation of cosmic distance duality relation, and parameterized the variation in $\alpha$ as a logarithmic function of redshift and encoded in the parameter $\gamma$ (defined in Eq.~\ref{eq:deltaalphagamma}). This logarithmic variation is  characteristic of runaway dilaton models~\cite{damour,damour1,martins}. 
 
 When we fit the $Y_{SZ}$ to $Y_X$ ratio for the SPT-SZ data to this model, our best-fit values are given by $\gamma = -0.046_{-0.1}^{+0.097}$ and $C = 0.95_{-0.11}^{+0.12}$. Therefore, our results  show no significant variation of the $\alpha$ with redshift, in accord with previous results using clusters. Similar to C19, our inferred value of $C$ also  indicates that our samples are well approximated by the isothermal temperature profile. A comparison of our results with previous results in literature are summarized in Table~\ref{tab:example_table}.

Finally,  similar to~\cite{galli}, we also search for a  dipole variation of $\alpha$  with the best-fit direction equal to the  value found using quasar-based searches~\cite{Webbspatial}.
The model we use to test for dipole variation is given in Eq.~\ref{eq:dipoleeq}. Our results show no such spatial  variation of $\alpha$, in agreement with  previous studies (cf. Table~\ref{tab:example_table_var}).

The first generation SZ surveys from the SPT and ACT telescopes have been superseded by SPTpol, SPT-3G~\cite{Benson14,Bleem19} and ACTpol~\cite{Thornton} respectively, and should detect about an order of magnitude more clusters upto redshift of 2.0. On the X-ray side, the recently launched eROSITA satellite should discover upto 100,000 clusters~\cite{erosita}. Therefore, further robust tests of both the temporal and spatial variations of $\alpha$ should soon be possible.

\section*{ACKNOWLEDGEMENT}
KB acknowledges Department of Science and Technology, Government of India for providing the financial support under DST-INSPIRE Fellowship program. \rthis{We are grateful to the anonymous referee for useful and constructive  feedback on the manuscript.}

%\input acknowledgement.tex   % input acknowledgement

%\bibsection{}
%\section*{REFERENCES}
\bibliography{ref}  

\begin{thebibliography}{102}
\expandafter\ifx\csname natexlab\endcsname\relax\def\natexlab#1{#1}\fi
\expandafter\ifx\csname bibnamefont\endcsname\relax
  \def\bibnamefont#1{#1}\fi
\expandafter\ifx\csname bibfnamefont\endcsname\relax
  \def\bibfnamefont#1{#1}\fi
\expandafter\ifx\csname citenamefont\endcsname\relax
  \def\citenamefont#1{#1}\fi
\expandafter\ifx\csname url\endcsname\relax
  \def\url#1{\texttt{#1}}\fi
\expandafter\ifx\csname urlprefix\endcsname\relax\def\urlprefix{URL }\fi
\providecommand{\bibinfo}[2]{#2}
\providecommand{\eprint}[2][]{\url{#2}}

\bibitem[{\citenamefont{{Dirac}}(1937)}]{Dirac}
\bibinfo{author}{\bibfnamefont{P.~A.~M.} \bibnamefont{{Dirac}}},
  \bibinfo{journal}{\nat} \textbf{\bibinfo{volume}{139}}, \bibinfo{pages}{323}
  (\bibinfo{year}{1937}).

\bibitem[{\citenamefont{{Uzan}}(2011)}]{Uzan}
\bibinfo{author}{\bibfnamefont{J.-P.} \bibnamefont{{Uzan}}},
  \bibinfo{journal}{Living Reviews in Relativity}
  \textbf{\bibinfo{volume}{14}}, \bibinfo{eid}{2} (\bibinfo{year}{2011}),
  \eprint{1009.5514}.

\bibitem[{\citenamefont{{Martins}}(2017)}]{Martinisreview}
\bibinfo{author}{\bibfnamefont{C.~J.~A.~P.} \bibnamefont{{Martins}}},
  \bibinfo{journal}{Reports on Progress in Physics}
  \textbf{\bibinfo{volume}{80}}, \bibinfo{eid}{126902} (\bibinfo{year}{2017}),
  \eprint{1709.02923}.

\bibitem[{\citenamefont{{Bekenstein}}(1982)}]{Bekenstein}
\bibinfo{author}{\bibfnamefont{J.~D.} \bibnamefont{{Bekenstein}}},
  \bibinfo{journal}{\prd} \textbf{\bibinfo{volume}{25}}, \bibinfo{pages}{1527}
  (\bibinfo{year}{1982}).

\bibitem[{\citenamefont{{Damour} and {Polyakov}}(1994)}]{Polyakov}
\bibinfo{author}{\bibfnamefont{T.}~\bibnamefont{{Damour}}} \bibnamefont{and}
  \bibinfo{author}{\bibfnamefont{A.~M.} \bibnamefont{{Polyakov}}},
  \bibinfo{journal}{Nuclear Physics B} \textbf{\bibinfo{volume}{423}},
  \bibinfo{pages}{532} (\bibinfo{year}{1994}), \eprint{hep-th/9401069}.

\bibitem[{\citenamefont{Sandvik et~al.}(2002)\citenamefont{Sandvik, Barrow, and
  Magueijo}}]{Barrow}
\bibinfo{author}{\bibfnamefont{H.~B.} \bibnamefont{Sandvik}},
  \bibinfo{author}{\bibfnamefont{J.~D.} \bibnamefont{Barrow}},
  \bibnamefont{and} \bibinfo{author}{\bibfnamefont{J.}~\bibnamefont{Magueijo}},
  \bibinfo{journal}{Phys. Rev. Lett.} \textbf{\bibinfo{volume}{88}},
  \bibinfo{pages}{031302} (\bibinfo{year}{2002}), \eprint{astro-ph/0107512}.

\bibitem[{\citenamefont{{Barrow} and {Lip}}(2012)}]{Barrow2}
\bibinfo{author}{\bibfnamefont{J.~D.} \bibnamefont{{Barrow}}} \bibnamefont{and}
  \bibinfo{author}{\bibfnamefont{S.~Z.~W.} \bibnamefont{{Lip}}},
  \bibinfo{journal}{\prd} \textbf{\bibinfo{volume}{85}}, \bibinfo{eid}{023514}
  (\bibinfo{year}{2012}), \eprint{1110.3120}.

\bibitem[{\citenamefont{{Webb} et~al.}(2001)\citenamefont{{Webb}, {Murphy},
  {Flambaum}, {Dzuba}, {Barrow}, {Churchill}, {Prochaska}, and {Wolfe}}}]{Webb}
\bibinfo{author}{\bibfnamefont{J.~K.} \bibnamefont{{Webb}}},
  \bibinfo{author}{\bibfnamefont{M.~T.} \bibnamefont{{Murphy}}},
  \bibinfo{author}{\bibfnamefont{V.~V.} \bibnamefont{{Flambaum}}},
  \bibinfo{author}{\bibfnamefont{V.~A.} \bibnamefont{{Dzuba}}},
  \bibinfo{author}{\bibfnamefont{J.~D.} \bibnamefont{{Barrow}}},
  \bibinfo{author}{\bibfnamefont{C.~W.} \bibnamefont{{Churchill}}},
  \bibinfo{author}{\bibfnamefont{J.~X.} \bibnamefont{{Prochaska}}},
  \bibnamefont{and} \bibinfo{author}{\bibfnamefont{A.~M.}
  \bibnamefont{{Wolfe}}}, \bibinfo{journal}{\prl}
  \textbf{\bibinfo{volume}{87}}, \bibinfo{eid}{091301} (\bibinfo{year}{2001}),
  \eprint{astro-ph/0012539}.

\bibitem[{\citenamefont{{Webb} et~al.}(2011)\citenamefont{{Webb}, {King},
  {Murphy}, {Flambaum}, {Carswell}, and {Bainbridge}}}]{Webbspatial}
\bibinfo{author}{\bibfnamefont{J.~K.} \bibnamefont{{Webb}}},
  \bibinfo{author}{\bibfnamefont{J.~A.} \bibnamefont{{King}}},
  \bibinfo{author}{\bibfnamefont{M.~T.} \bibnamefont{{Murphy}}},
  \bibinfo{author}{\bibfnamefont{V.~V.} \bibnamefont{{Flambaum}}},
  \bibinfo{author}{\bibfnamefont{R.~F.} \bibnamefont{{Carswell}}},
  \bibnamefont{and} \bibinfo{author}{\bibfnamefont{M.~B.}
  \bibnamefont{{Bainbridge}}}, \bibinfo{journal}{\prl}
  \textbf{\bibinfo{volume}{107}}, \bibinfo{eid}{191101} (\bibinfo{year}{2011}),
  \eprint{1008.3907}.

\bibitem[{\citenamefont{{King} et~al.}(2012)\citenamefont{{King}, {Webb},
  {Murphy}, {Flambaum}, {Carswell}, {Bainbridge}, {Wilczynska}, and
  {Koch}}}]{King}
\bibinfo{author}{\bibfnamefont{J.~A.} \bibnamefont{{King}}},
  \bibinfo{author}{\bibfnamefont{J.~K.} \bibnamefont{{Webb}}},
  \bibinfo{author}{\bibfnamefont{M.~T.} \bibnamefont{{Murphy}}},
  \bibinfo{author}{\bibfnamefont{V.~V.} \bibnamefont{{Flambaum}}},
  \bibinfo{author}{\bibfnamefont{R.~F.} \bibnamefont{{Carswell}}},
  \bibinfo{author}{\bibfnamefont{M.~B.} \bibnamefont{{Bainbridge}}},
  \bibinfo{author}{\bibfnamefont{M.~R.} \bibnamefont{{Wilczynska}}},
  \bibnamefont{and} \bibinfo{author}{\bibfnamefont{F.~E.}
  \bibnamefont{{Koch}}}, \bibinfo{journal}{\mnras}
  \textbf{\bibinfo{volume}{422}}, \bibinfo{pages}{3370} (\bibinfo{year}{2012}),
  \eprint{1202.4758}.

\bibitem[{\citenamefont{{Srianand} et~al.}(2004)\citenamefont{{Srianand},
  {Chand}, {Petitjean}, and {Aracil}}}]{Srianand}
\bibinfo{author}{\bibfnamefont{R.}~\bibnamefont{{Srianand}}},
  \bibinfo{author}{\bibfnamefont{H.}~\bibnamefont{{Chand}}},
  \bibinfo{author}{\bibfnamefont{P.}~\bibnamefont{{Petitjean}}},
  \bibnamefont{and} \bibinfo{author}{\bibfnamefont{B.}~\bibnamefont{{Aracil}}},
  \bibinfo{journal}{\prl} \textbf{\bibinfo{volume}{92}}, \bibinfo{eid}{121302}
  (\bibinfo{year}{2004}), \eprint{astro-ph/0402177}.

\bibitem[{\citenamefont{{Whitmore} and {Murphy}}(2015)}]{Whitmore15}
\bibinfo{author}{\bibfnamefont{J.~B.} \bibnamefont{{Whitmore}}}
  \bibnamefont{and} \bibinfo{author}{\bibfnamefont{M.~T.}
  \bibnamefont{{Murphy}}}, \bibinfo{journal}{\mnras}
  \textbf{\bibinfo{volume}{447}}, \bibinfo{pages}{446} (\bibinfo{year}{2015}),
  \eprint{1409.4467}.

\bibitem[{\citenamefont{{Murphy} et~al.}(2016)\citenamefont{{Murphy}, {Malec},
  and {Prochaska}}}]{Prochaska16}
\bibinfo{author}{\bibfnamefont{M.~T.} \bibnamefont{{Murphy}}},
  \bibinfo{author}{\bibfnamefont{A.~L.} \bibnamefont{{Malec}}},
  \bibnamefont{and} \bibinfo{author}{\bibfnamefont{J.~X.}
  \bibnamefont{{Prochaska}}}, \bibinfo{journal}{\mnras}
  \textbf{\bibinfo{volume}{461}}, \bibinfo{pages}{2461} (\bibinfo{year}{2016}),
  \eprint{1606.06293}.

\bibitem[{\citenamefont{{Murphy} et~al.}(2017)\citenamefont{{Murphy}, {Malec},
  and {Prochaska}}}]{Prochaska17}
\bibinfo{author}{\bibfnamefont{M.~T.} \bibnamefont{{Murphy}}},
  \bibinfo{author}{\bibfnamefont{A.~L.} \bibnamefont{{Malec}}},
  \bibnamefont{and} \bibinfo{author}{\bibfnamefont{J.~X.}
  \bibnamefont{{Prochaska}}}, \bibinfo{journal}{\mnras}
  \textbf{\bibinfo{volume}{464}}, \bibinfo{pages}{2609} (\bibinfo{year}{2017}).

\bibitem[{\citenamefont{{Verde} et~al.}(2019)\citenamefont{{Verde}, {Treu}, and
  {Riess}}}]{Riess}
\bibinfo{author}{\bibfnamefont{L.}~\bibnamefont{{Verde}}},
  \bibinfo{author}{\bibfnamefont{T.}~\bibnamefont{{Treu}}}, \bibnamefont{and}
  \bibinfo{author}{\bibfnamefont{A.~G.} \bibnamefont{{Riess}}},
  \bibinfo{journal}{Nature Astronomy} \textbf{\bibinfo{volume}{3}},
  \bibinfo{pages}{891} (\bibinfo{year}{2019}), \eprint{1907.10625}.

\bibitem[{\citenamefont{Bethapudi and Desai}(2017)}]{Bethapudi}
\bibinfo{author}{\bibfnamefont{S.}~\bibnamefont{Bethapudi}} \bibnamefont{and}
  \bibinfo{author}{\bibfnamefont{S.}~\bibnamefont{Desai}},
  \bibinfo{journal}{Eur. Phys. J. Plus} \textbf{\bibinfo{volume}{132}},
  \bibinfo{pages}{78} (\bibinfo{year}{2017}), \eprint{1701.01789}.

\bibitem[{\citenamefont{{Knox} and {Millea}}(2020)}]{Knox}
\bibinfo{author}{\bibfnamefont{L.}~\bibnamefont{{Knox}}} \bibnamefont{and}
  \bibinfo{author}{\bibfnamefont{M.}~\bibnamefont{{Millea}}},
  \bibinfo{journal}{\prd} \textbf{\bibinfo{volume}{101}}, \bibinfo{eid}{043533}
  (\bibinfo{year}{2020}), \eprint{1908.03663}.

\bibitem[{\citenamefont{{Hart} and {Chluba}}(2020)}]{Hart2}
\bibinfo{author}{\bibfnamefont{L.}~\bibnamefont{{Hart}}} \bibnamefont{and}
  \bibinfo{author}{\bibfnamefont{J.}~\bibnamefont{{Chluba}}},
  \bibinfo{journal}{\mnras} \textbf{\bibinfo{volume}{493}},
  \bibinfo{pages}{3255} (\bibinfo{year}{2020}), \eprint{1912.03986}.

\bibitem[{\citenamefont{{Menegoni} et~al.}(2012)\citenamefont{{Menegoni},
  {Archidiacono}, {Calabrese}, {Galli}, {Martins}, and
  {Melchiorri}}}]{Menegoni}
\bibinfo{author}{\bibfnamefont{E.}~\bibnamefont{{Menegoni}}},
  \bibinfo{author}{\bibfnamefont{M.}~\bibnamefont{{Archidiacono}}},
  \bibinfo{author}{\bibfnamefont{E.}~\bibnamefont{{Calabrese}}},
  \bibinfo{author}{\bibfnamefont{S.}~\bibnamefont{{Galli}}},
  \bibinfo{author}{\bibfnamefont{C.~J.~A.~P.} \bibnamefont{{Martins}}},
  \bibnamefont{and}
  \bibinfo{author}{\bibfnamefont{A.}~\bibnamefont{{Melchiorri}}},
  \bibinfo{journal}{\prd} \textbf{\bibinfo{volume}{85}}, \bibinfo{eid}{107301}
  (\bibinfo{year}{2012}), \eprint{1202.1476}.

\bibitem[{\citenamefont{Ade et~al.}(2015)}]{Planck14}
\bibinfo{author}{\bibfnamefont{P.}~\bibnamefont{Ade}} \bibnamefont{et~al.}
  (\bibinfo{collaboration}{Planck}), \bibinfo{journal}{Astron. Astrophys.}
  \textbf{\bibinfo{volume}{580}}, \bibinfo{pages}{A22} (\bibinfo{year}{2015}),
  \eprint{1406.7482}.

\bibitem[{\citenamefont{{Hart} and {Chluba}}(2018)}]{Hart}
\bibinfo{author}{\bibfnamefont{L.}~\bibnamefont{{Hart}}} \bibnamefont{and}
  \bibinfo{author}{\bibfnamefont{J.}~\bibnamefont{{Chluba}}},
  \bibinfo{journal}{\mnras} \textbf{\bibinfo{volume}{474}},
  \bibinfo{pages}{1850} (\bibinfo{year}{2018}), \eprint{1705.03925}.

\bibitem[{\citenamefont{{Smith} et~al.}(2019)\citenamefont{{Smith}, {Grin},
  {Robinson}, and {Qi}}}]{smith19}
\bibinfo{author}{\bibfnamefont{T.~L.} \bibnamefont{{Smith}}},
  \bibinfo{author}{\bibfnamefont{D.}~\bibnamefont{{Grin}}},
  \bibinfo{author}{\bibfnamefont{D.}~\bibnamefont{{Robinson}}},
  \bibnamefont{and} \bibinfo{author}{\bibfnamefont{D.}~\bibnamefont{{Qi}}},
  \bibinfo{journal}{\prd} \textbf{\bibinfo{volume}{99}}, \bibinfo{eid}{043531}
  (\bibinfo{year}{2019}), \eprint{1808.07486}.

\bibitem[{\citenamefont{{Clara} and {Martins}}(2020)}]{Ciara}
\bibinfo{author}{\bibfnamefont{M.~T.} \bibnamefont{{Clara}}} \bibnamefont{and}
  \bibinfo{author}{\bibfnamefont{C.~J.~A.~P.} \bibnamefont{{Martins}}},
  \bibinfo{journal}{\aap} \textbf{\bibinfo{volume}{633}}, \bibinfo{eid}{L11}
  (\bibinfo{year}{2020}), \eprint{2001.01787}.

\bibitem[{\citenamefont{{Hees} et~al.}(2020)\citenamefont{{Hees}, {Do},
  {Roberts}, {Ghez}, {Nishiyama}, {Bentley}, {Gautam}, {Jia}, {Kara}, {Lu}
  et~al.}}]{Ghez}
\bibinfo{author}{\bibfnamefont{A.}~\bibnamefont{{Hees}}},
  \bibinfo{author}{\bibfnamefont{T.}~\bibnamefont{{Do}}},
  \bibinfo{author}{\bibfnamefont{B.~M.} \bibnamefont{{Roberts}}},
  \bibinfo{author}{\bibfnamefont{A.~M.} \bibnamefont{{Ghez}}},
  \bibinfo{author}{\bibfnamefont{S.}~\bibnamefont{{Nishiyama}}},
  \bibinfo{author}{\bibfnamefont{R.~O.} \bibnamefont{{Bentley}}},
  \bibinfo{author}{\bibfnamefont{A.~K.} \bibnamefont{{Gautam}}},
  \bibinfo{author}{\bibfnamefont{S.}~\bibnamefont{{Jia}}},
  \bibinfo{author}{\bibfnamefont{T.}~\bibnamefont{{Kara}}},
  \bibinfo{author}{\bibfnamefont{J.~R.} \bibnamefont{{Lu}}},
  \bibnamefont{et~al.}, \bibinfo{journal}{\prl} \textbf{\bibinfo{volume}{124}},
  \bibinfo{eid}{081101} (\bibinfo{year}{2020}), \eprint{2002.11567}.

\bibitem[{\citenamefont{{Berengut} et~al.}(2013)\citenamefont{{Berengut},
  {Flambaum}, {Ong}, {Webb}, {Barrow}, {Barstow}, {Preval}, and
  {Holberg}}}]{whitedwarf}
\bibinfo{author}{\bibfnamefont{J.~C.} \bibnamefont{{Berengut}}},
  \bibinfo{author}{\bibfnamefont{V.~V.} \bibnamefont{{Flambaum}}},
  \bibinfo{author}{\bibfnamefont{A.}~\bibnamefont{{Ong}}},
  \bibinfo{author}{\bibfnamefont{J.~K.} \bibnamefont{{Webb}}},
  \bibinfo{author}{\bibfnamefont{J.~D.} \bibnamefont{{Barrow}}},
  \bibinfo{author}{\bibfnamefont{M.~A.} \bibnamefont{{Barstow}}},
  \bibinfo{author}{\bibfnamefont{S.~P.} \bibnamefont{{Preval}}},
  \bibnamefont{and} \bibinfo{author}{\bibfnamefont{J.~B.}
  \bibnamefont{{Holberg}}}, \bibinfo{journal}{\prl}
  \textbf{\bibinfo{volume}{111}}, \bibinfo{eid}{010801} (\bibinfo{year}{2013}),
  \eprint{1305.1337}.

\bibitem[{\citenamefont{{Cola{\c{c}}o}
  et~al.}(2020)\citenamefont{{Cola{\c{c}}o}, {Holanda}, and {Silva}}}]{Cola20}
\bibinfo{author}{\bibfnamefont{L.~R.} \bibnamefont{{Cola{\c{c}}o}}},
  \bibinfo{author}{\bibfnamefont{R.~F.~L.} \bibnamefont{{Holanda}}},
  \bibnamefont{and} \bibinfo{author}{\bibfnamefont{R.}~\bibnamefont{{Silva}}},
  \bibinfo{journal}{arXiv e-prints} \bibinfo{eid}{arXiv:2004.08484}
  (\bibinfo{year}{2020}), \eprint{2004.08484}.

\bibitem[{\citenamefont{{Damour} and {Dyson}}(1996)}]{Dyson}
\bibinfo{author}{\bibfnamefont{T.}~\bibnamefont{{Damour}}} \bibnamefont{and}
  \bibinfo{author}{\bibfnamefont{F.}~\bibnamefont{{Dyson}}},
  \bibinfo{journal}{Nuclear Physics B} \textbf{\bibinfo{volume}{480}},
  \bibinfo{pages}{37} (\bibinfo{year}{1996}), \eprint{hep-ph/9606486}.

\bibitem[{\citenamefont{{Godun} et~al.}(2014)\citenamefont{{Godun},
  {Nisbet-Jones}, {Jones}, {King}, {Johnson}, {Margolis}, {Szymaniec}, {Lea},
  {Bongs}, and {Gill}}}]{Godun}
\bibinfo{author}{\bibfnamefont{R.~M.} \bibnamefont{{Godun}}},
  \bibinfo{author}{\bibfnamefont{P.~B.~R.} \bibnamefont{{Nisbet-Jones}}},
  \bibinfo{author}{\bibfnamefont{J.~M.} \bibnamefont{{Jones}}},
  \bibinfo{author}{\bibfnamefont{S.~A.} \bibnamefont{{King}}},
  \bibinfo{author}{\bibfnamefont{L.~A.~M.} \bibnamefont{{Johnson}}},
  \bibinfo{author}{\bibfnamefont{H.~S.} \bibnamefont{{Margolis}}},
  \bibinfo{author}{\bibfnamefont{K.}~\bibnamefont{{Szymaniec}}},
  \bibinfo{author}{\bibfnamefont{S.~N.} \bibnamefont{{Lea}}},
  \bibinfo{author}{\bibfnamefont{K.}~\bibnamefont{{Bongs}}}, \bibnamefont{and}
  \bibinfo{author}{\bibfnamefont{P.}~\bibnamefont{{Gill}}},
  \bibinfo{journal}{\prl} \textbf{\bibinfo{volume}{113}}, \bibinfo{eid}{210801}
  (\bibinfo{year}{2014}), \eprint{1407.0164}.

\bibitem[{\citenamefont{{Allen} et~al.}(2011)\citenamefont{{Allen}, {Evrard},
  and {Mantz}}}]{Allen}
\bibinfo{author}{\bibfnamefont{S.~W.} \bibnamefont{{Allen}}},
  \bibinfo{author}{\bibfnamefont{A.~E.} \bibnamefont{{Evrard}}},
  \bibnamefont{and} \bibinfo{author}{\bibfnamefont{A.~B.}
  \bibnamefont{{Mantz}}}, \bibinfo{journal}{\araa}
  \textbf{\bibinfo{volume}{49}}, \bibinfo{pages}{409} (\bibinfo{year}{2011}),
  \eprint{1103.4829}.

\bibitem[{\citenamefont{{Vikhlinin} et~al.}(2014)\citenamefont{{Vikhlinin},
  {Kravtsov}, {Markevich}, {Sunyaev}, and {Churazov}}}]{Vikhlininrev}
\bibinfo{author}{\bibfnamefont{A.~A.} \bibnamefont{{Vikhlinin}}},
  \bibinfo{author}{\bibfnamefont{A.~V.} \bibnamefont{{Kravtsov}}},
  \bibinfo{author}{\bibfnamefont{M.~L.} \bibnamefont{{Markevich}}},
  \bibinfo{author}{\bibfnamefont{R.~A.} \bibnamefont{{Sunyaev}}},
  \bibnamefont{and} \bibinfo{author}{\bibfnamefont{E.~M.}
  \bibnamefont{{Churazov}}}, \bibinfo{journal}{Physics Uspekhi}
  \textbf{\bibinfo{volume}{57}}, \bibinfo{eid}{317-341} (\bibinfo{year}{2014}).

\bibitem[{\citenamefont{{Desai} and {Gupta}}(2020)}]{Desai}
\bibinfo{author}{\bibfnamefont{S.}~\bibnamefont{{Desai}}} \bibnamefont{and}
  \bibinfo{author}{\bibfnamefont{S.}~\bibnamefont{{Gupta}}}
  (\bibinfo{year}{2020}), vol. \bibinfo{volume}{1468} of
  \emph{\bibinfo{series}{Journal of Physics Conference Series}}, p.
  \bibinfo{pages}{012003}, \eprint{1912.05117}.

\bibitem[{\citenamefont{{Galli}}(2013)}]{galli}
\bibinfo{author}{\bibfnamefont{S.}~\bibnamefont{{Galli}}},
  \bibinfo{journal}{\prd} \textbf{\bibinfo{volume}{87}}, \bibinfo{eid}{123516}
  (\bibinfo{year}{2013}), \eprint{1212.1075}.

\bibitem[{\citenamefont{{Planck Collaboration}
  et~al.}(2011)\citenamefont{{Planck Collaboration}, {Ade}, {Aghanim},
  {Arnaud}, {Ashdown}, {Aumont}, {Baccigalupi}, {Balbi}, {Banday}, {Barreiro}
  et~al.}}]{planck}
\bibinfo{author}{\bibnamefont{{Planck Collaboration}}},
  \bibinfo{author}{\bibfnamefont{P.~A.~R.} \bibnamefont{{Ade}}},
  \bibinfo{author}{\bibfnamefont{N.}~\bibnamefont{{Aghanim}}},
  \bibinfo{author}{\bibfnamefont{M.}~\bibnamefont{{Arnaud}}},
  \bibinfo{author}{\bibfnamefont{M.}~\bibnamefont{{Ashdown}}},
  \bibinfo{author}{\bibfnamefont{J.}~\bibnamefont{{Aumont}}},
  \bibinfo{author}{\bibfnamefont{C.}~\bibnamefont{{Baccigalupi}}},
  \bibinfo{author}{\bibfnamefont{A.}~\bibnamefont{{Balbi}}},
  \bibinfo{author}{\bibfnamefont{A.~J.} \bibnamefont{{Banday}}},
  \bibinfo{author}{\bibfnamefont{R.~B.} \bibnamefont{{Barreiro}}},
  \bibnamefont{et~al.}, \bibinfo{journal}{\aap} \textbf{\bibinfo{volume}{536}},
  \bibinfo{eid}{A11} (\bibinfo{year}{2011}), \eprint{1101.2026}.

\bibitem[{\citenamefont{{Holanda}
  et~al.}(2016{\natexlab{a}})\citenamefont{{Holanda}, {Landau}, {Alcaniz},
  {S{\'a}nchez G.}, and {Busti}}}]{hola1}
\bibinfo{author}{\bibfnamefont{R.~F.~L.} \bibnamefont{{Holanda}}},
  \bibinfo{author}{\bibfnamefont{S.~J.} \bibnamefont{{Landau}}},
  \bibinfo{author}{\bibfnamefont{J.~S.} \bibnamefont{{Alcaniz}}},
  \bibinfo{author}{\bibfnamefont{I.~E.} \bibnamefont{{S{\'a}nchez G.}}},
  \bibnamefont{and} \bibinfo{author}{\bibfnamefont{V.~C.}
  \bibnamefont{{Busti}}}, \bibinfo{journal}{\jcap}
  \textbf{\bibinfo{volume}{2016}}, \bibinfo{eid}{047}
  (\bibinfo{year}{2016}{\natexlab{a}}), \eprint{1510.07240}.

\bibitem[{\citenamefont{{Etherington}}(1933)}]{Etherington}
\bibinfo{author}{\bibfnamefont{I.~M.~H.} \bibnamefont{{Etherington}}},
  \bibinfo{journal}{Philosophical Magazine} \textbf{\bibinfo{volume}{15}},
  \bibinfo{pages}{761} (\bibinfo{year}{1933}).

\bibitem[{\citenamefont{{LaRoque} et~al.}(2006)\citenamefont{{LaRoque},
  {Bonamente}, {Carlstrom}, {Joy}, {Nagai}, {Reese}, and {Dawson}}}]{Roque}
\bibinfo{author}{\bibfnamefont{S.~J.} \bibnamefont{{LaRoque}}},
  \bibinfo{author}{\bibfnamefont{M.}~\bibnamefont{{Bonamente}}},
  \bibinfo{author}{\bibfnamefont{J.~E.} \bibnamefont{{Carlstrom}}},
  \bibinfo{author}{\bibfnamefont{M.~K.} \bibnamefont{{Joy}}},
  \bibinfo{author}{\bibfnamefont{D.}~\bibnamefont{{Nagai}}},
  \bibinfo{author}{\bibfnamefont{E.~D.} \bibnamefont{{Reese}}},
  \bibnamefont{and} \bibinfo{author}{\bibfnamefont{K.~S.}
  \bibnamefont{{Dawson}}}, \bibinfo{journal}{\apj}
  \textbf{\bibinfo{volume}{652}}, \bibinfo{pages}{917} (\bibinfo{year}{2006}),
  \eprint{astro-ph/0604039}.

\bibitem[{\citenamefont{{Holanda}
  et~al.}(2016{\natexlab{b}})\citenamefont{{Holanda}, {Busti}, {Cola{\c{c}}o},
  {Alcaniz}, and {Landau}}}]{hola2}
\bibinfo{author}{\bibfnamefont{R.~F.~L.} \bibnamefont{{Holanda}}},
  \bibinfo{author}{\bibfnamefont{V.~C.} \bibnamefont{{Busti}}},
  \bibinfo{author}{\bibfnamefont{L.~R.} \bibnamefont{{Cola{\c{c}}o}}},
  \bibinfo{author}{\bibfnamefont{J.~S.} \bibnamefont{{Alcaniz}}},
  \bibnamefont{and} \bibinfo{author}{\bibfnamefont{S.~J.}
  \bibnamefont{{Landau}}}, \bibinfo{journal}{\jcap}
  \textbf{\bibinfo{volume}{2016}}, \bibinfo{eid}{055}
  (\bibinfo{year}{2016}{\natexlab{b}}), \eprint{1605.02578}.

\bibitem[{\citenamefont{{Holanda} et~al.}(2017)\citenamefont{{Holanda},
  {Cola{\c{c}}o}, {Gon{\c{c}}alves}, and {Alcaniz}}}]{hola3}
\bibinfo{author}{\bibfnamefont{R.~F.~L.} \bibnamefont{{Holanda}}},
  \bibinfo{author}{\bibfnamefont{L.~R.} \bibnamefont{{Cola{\c{c}}o}}},
  \bibinfo{author}{\bibfnamefont{R.~S.} \bibnamefont{{Gon{\c{c}}alves}}},
  \bibnamefont{and} \bibinfo{author}{\bibfnamefont{J.~S.}
  \bibnamefont{{Alcaniz}}}, \bibinfo{journal}{Physics Letters B}
  \textbf{\bibinfo{volume}{767}}, \bibinfo{pages}{188} (\bibinfo{year}{2017}),
  \eprint{1701.07250}.

\bibitem[{\citenamefont{{de Martino} et~al.}(2016)\citenamefont{{de Martino},
  {Martins}, {Ebeling}, and {Kocevski}}}]{martino16D}
\bibinfo{author}{\bibfnamefont{I.}~\bibnamefont{{de Martino}}},
  \bibinfo{author}{\bibfnamefont{C.~J.~A.~P.} \bibnamefont{{Martins}}},
  \bibinfo{author}{\bibfnamefont{H.}~\bibnamefont{{Ebeling}}},
  \bibnamefont{and}
  \bibinfo{author}{\bibfnamefont{D.}~\bibnamefont{{Kocevski}}},
  \bibinfo{journal}{Universe} \textbf{\bibinfo{volume}{2}}, \bibinfo{eid}{34}
  (\bibinfo{year}{2016}), \eprint{1612.06739}.

\bibitem[{\citenamefont{{Cola{\c{c}}o}
  et~al.}(2019)\citenamefont{{Cola{\c{c}}o}, {Holanda}, {Silva}, and
  {Alcaniz}}}]{colaco}
\bibinfo{author}{\bibfnamefont{L.~R.} \bibnamefont{{Cola{\c{c}}o}}},
  \bibinfo{author}{\bibfnamefont{R.~F.~L.} \bibnamefont{{Holanda}}},
  \bibinfo{author}{\bibfnamefont{R.}~\bibnamefont{{Silva}}}, \bibnamefont{and}
  \bibinfo{author}{\bibfnamefont{J.~S.} \bibnamefont{{Alcaniz}}},
  \bibinfo{journal}{\jcap} \textbf{\bibinfo{volume}{2019}}, \bibinfo{eid}{014}
  (\bibinfo{year}{2019}), \eprint{1901.10947}.

\bibitem[{\citenamefont{{Damour}
  et~al.}(2002{\natexlab{a}})\citenamefont{{Damour}, {Piazza}, and
  {Veneziano}}}]{damour}
\bibinfo{author}{\bibfnamefont{T.}~\bibnamefont{{Damour}}},
  \bibinfo{author}{\bibfnamefont{F.}~\bibnamefont{{Piazza}}}, \bibnamefont{and}
  \bibinfo{author}{\bibfnamefont{G.}~\bibnamefont{{Veneziano}}},
  \bibinfo{journal}{\prd} \textbf{\bibinfo{volume}{66}}, \bibinfo{eid}{046007}
  (\bibinfo{year}{2002}{\natexlab{a}}), \eprint{hep-th/0205111}.

\bibitem[{\citenamefont{{Damour}
  et~al.}(2002{\natexlab{b}})\citenamefont{{Damour}, {Piazza}, and
  {Veneziano}}}]{damour1}
\bibinfo{author}{\bibfnamefont{T.}~\bibnamefont{{Damour}}},
  \bibinfo{author}{\bibfnamefont{F.}~\bibnamefont{{Piazza}}}, \bibnamefont{and}
  \bibinfo{author}{\bibfnamefont{G.}~\bibnamefont{{Veneziano}}},
  \bibinfo{journal}{\prl} \textbf{\bibinfo{volume}{89}}, \bibinfo{eid}{081601}
  (\bibinfo{year}{2002}{\natexlab{b}}), \eprint{gr-qc/0204094}.

\bibitem[{\citenamefont{{Martins} et~al.}(2015)\citenamefont{{Martins},
  {Vielzeuf}, {Martinelli}, {Calabrese}, and {Pandolfi}}}]{martins}
\bibinfo{author}{\bibfnamefont{C.~J.~A.~P.} \bibnamefont{{Martins}}},
  \bibinfo{author}{\bibfnamefont{P.~E.} \bibnamefont{{Vielzeuf}}},
  \bibinfo{author}{\bibfnamefont{M.}~\bibnamefont{{Martinelli}}},
  \bibinfo{author}{\bibfnamefont{E.}~\bibnamefont{{Calabrese}}},
  \bibnamefont{and}
  \bibinfo{author}{\bibfnamefont{S.}~\bibnamefont{{Pandolfi}}},
  \bibinfo{journal}{Physics Letters B} \textbf{\bibinfo{volume}{743}},
  \bibinfo{pages}{377} (\bibinfo{year}{2015}), \eprint{1503.05068}.

\bibitem[{\citenamefont{{Planelles} et~al.}(2017)\citenamefont{{Planelles},
  {Fabjan}, {Borgani}, {Murante}, {Rasia}, {Biffi}, {Truong},
  {Ragone-Figueroa}, {Granato}, {Dolag} et~al.}}]{Planelles17}
\bibinfo{author}{\bibfnamefont{S.}~\bibnamefont{{Planelles}}},
  \bibinfo{author}{\bibfnamefont{D.}~\bibnamefont{{Fabjan}}},
  \bibinfo{author}{\bibfnamefont{S.}~\bibnamefont{{Borgani}}},
  \bibinfo{author}{\bibfnamefont{G.}~\bibnamefont{{Murante}}},
  \bibinfo{author}{\bibfnamefont{E.}~\bibnamefont{{Rasia}}},
  \bibinfo{author}{\bibfnamefont{V.}~\bibnamefont{{Biffi}}},
  \bibinfo{author}{\bibfnamefont{N.}~\bibnamefont{{Truong}}},
  \bibinfo{author}{\bibfnamefont{C.}~\bibnamefont{{Ragone-Figueroa}}},
  \bibinfo{author}{\bibfnamefont{G.~L.} \bibnamefont{{Granato}}},
  \bibinfo{author}{\bibfnamefont{K.}~\bibnamefont{{Dolag}}},
  \bibnamefont{et~al.}, \bibinfo{journal}{\mnras}
  \textbf{\bibinfo{volume}{467}}, \bibinfo{pages}{3827} (\bibinfo{year}{2017}),
  \eprint{1612.07260}.

\bibitem[{\citenamefont{{Bonamente} et~al.}(2008)\citenamefont{{Bonamente},
  {Joy}, {LaRoque}, {Carlstrom}, {Nagai}, and {Marrone}}}]{Bonamente}
\bibinfo{author}{\bibfnamefont{M.}~\bibnamefont{{Bonamente}}},
  \bibinfo{author}{\bibfnamefont{M.}~\bibnamefont{{Joy}}},
  \bibinfo{author}{\bibfnamefont{S.~J.} \bibnamefont{{LaRoque}}},
  \bibinfo{author}{\bibfnamefont{J.~E.} \bibnamefont{{Carlstrom}}},
  \bibinfo{author}{\bibfnamefont{D.}~\bibnamefont{{Nagai}}}, \bibnamefont{and}
  \bibinfo{author}{\bibfnamefont{D.~P.} \bibnamefont{{Marrone}}},
  \bibinfo{journal}{\apj} \textbf{\bibinfo{volume}{675}}, \bibinfo{pages}{106}
  (\bibinfo{year}{2008}), \eprint{0708.0815}.

\bibitem[{\citenamefont{{Andersson} et~al.}(2011)\citenamefont{{Andersson},
  {Benson}, {Ade}, {Aird}, {Armstrong}, {Bautz}, {Bleem}, {Brodwin},
  {Carlstrom}, {Chang} et~al.}}]{Andersson10}
\bibinfo{author}{\bibfnamefont{K.}~\bibnamefont{{Andersson}}},
  \bibinfo{author}{\bibfnamefont{B.~A.} \bibnamefont{{Benson}}},
  \bibinfo{author}{\bibfnamefont{P.~A.~R.} \bibnamefont{{Ade}}},
  \bibinfo{author}{\bibfnamefont{K.~A.} \bibnamefont{{Aird}}},
  \bibinfo{author}{\bibfnamefont{B.}~\bibnamefont{{Armstrong}}},
  \bibinfo{author}{\bibfnamefont{M.}~\bibnamefont{{Bautz}}},
  \bibinfo{author}{\bibfnamefont{L.~E.} \bibnamefont{{Bleem}}},
  \bibinfo{author}{\bibfnamefont{M.}~\bibnamefont{{Brodwin}}},
  \bibinfo{author}{\bibfnamefont{J.~E.} \bibnamefont{{Carlstrom}}},
  \bibinfo{author}{\bibfnamefont{C.~L.} \bibnamefont{{Chang}}},
  \bibnamefont{et~al.}, \bibinfo{journal}{\apj} \textbf{\bibinfo{volume}{738}},
  \bibinfo{eid}{48} (\bibinfo{year}{2011}), \eprint{1006.3068}.

\bibitem[{\citenamefont{{Bonamente} et~al.}(2012)\citenamefont{{Bonamente},
  {Hasler}, {Bulbul}, {Carlstrom}, {Culverhouse}, {Gralla}, {Greer}, {Hawkins},
  {Hennessy}, {Joy} et~al.}}]{Bonamente12}
\bibinfo{author}{\bibfnamefont{M.}~\bibnamefont{{Bonamente}}},
  \bibinfo{author}{\bibfnamefont{N.}~\bibnamefont{{Hasler}}},
  \bibinfo{author}{\bibfnamefont{E.}~\bibnamefont{{Bulbul}}},
  \bibinfo{author}{\bibfnamefont{J.~E.} \bibnamefont{{Carlstrom}}},
  \bibinfo{author}{\bibfnamefont{T.~L.} \bibnamefont{{Culverhouse}}},
  \bibinfo{author}{\bibfnamefont{M.}~\bibnamefont{{Gralla}}},
  \bibinfo{author}{\bibfnamefont{C.}~\bibnamefont{{Greer}}},
  \bibinfo{author}{\bibfnamefont{D.}~\bibnamefont{{Hawkins}}},
  \bibinfo{author}{\bibfnamefont{R.}~\bibnamefont{{Hennessy}}},
  \bibinfo{author}{\bibfnamefont{M.}~\bibnamefont{{Joy}}},
  \bibnamefont{et~al.}, \bibinfo{journal}{New Journal of Physics}
  \textbf{\bibinfo{volume}{14}}, \bibinfo{eid}{025010} (\bibinfo{year}{2012}),
  \eprint{1112.1599}.

\bibitem[{\citenamefont{{Rozo} et~al.}(2012)\citenamefont{{Rozo}, {Vikhlinin},
  and {More}}}]{More}
\bibinfo{author}{\bibfnamefont{E.}~\bibnamefont{{Rozo}}},
  \bibinfo{author}{\bibfnamefont{A.}~\bibnamefont{{Vikhlinin}}},
  \bibnamefont{and} \bibinfo{author}{\bibfnamefont{S.}~\bibnamefont{{More}}},
  \bibinfo{journal}{\apj} \textbf{\bibinfo{volume}{760}}, \bibinfo{eid}{67}
  (\bibinfo{year}{2012}), \eprint{1202.2150}.

\bibitem[{\citenamefont{{Rozo} et~al.}(2014{\natexlab{a}})\citenamefont{{Rozo},
  {Evrard}, {Rykoff}, and {Bartlett}}}]{Rozo}
\bibinfo{author}{\bibfnamefont{E.}~\bibnamefont{{Rozo}}},
  \bibinfo{author}{\bibfnamefont{A.~E.} \bibnamefont{{Evrard}}},
  \bibinfo{author}{\bibfnamefont{E.~S.} \bibnamefont{{Rykoff}}},
  \bibnamefont{and} \bibinfo{author}{\bibfnamefont{J.~G.}
  \bibnamefont{{Bartlett}}}, \bibinfo{journal}{\mnras}
  \textbf{\bibinfo{volume}{438}}, \bibinfo{pages}{62}
  (\bibinfo{year}{2014}{\natexlab{a}}), \eprint{1204.6292}.

\bibitem[{\citenamefont{{Rozo} et~al.}(2014{\natexlab{b}})\citenamefont{{Rozo},
  {Bartlett}, {Evrard}, and {Rykoff}}}]{Rozo2}
\bibinfo{author}{\bibfnamefont{E.}~\bibnamefont{{Rozo}}},
  \bibinfo{author}{\bibfnamefont{J.~G.} \bibnamefont{{Bartlett}}},
  \bibinfo{author}{\bibfnamefont{A.~E.} \bibnamefont{{Evrard}}},
  \bibnamefont{and} \bibinfo{author}{\bibfnamefont{E.~S.}
  \bibnamefont{{Rykoff}}}, \bibinfo{journal}{\mnras}
  \textbf{\bibinfo{volume}{438}}, \bibinfo{pages}{78}
  (\bibinfo{year}{2014}{\natexlab{b}}), \eprint{1204.6305}.

\bibitem[{\citenamefont{{Liu} et~al.}(2015)\citenamefont{{Liu}, {Mohr}, {Saro},
  {Aird}, {Ashby}, {Bautz}, {Bayliss}, {Benson}, {Bleem}, {Bocquet}
  et~al.}}]{Liu}
\bibinfo{author}{\bibfnamefont{J.}~\bibnamefont{{Liu}}},
  \bibinfo{author}{\bibfnamefont{J.}~\bibnamefont{{Mohr}}},
  \bibinfo{author}{\bibfnamefont{A.}~\bibnamefont{{Saro}}},
  \bibinfo{author}{\bibfnamefont{K.~A.} \bibnamefont{{Aird}}},
  \bibinfo{author}{\bibfnamefont{M.~L.~N.} \bibnamefont{{Ashby}}},
  \bibinfo{author}{\bibfnamefont{M.}~\bibnamefont{{Bautz}}},
  \bibinfo{author}{\bibfnamefont{M.}~\bibnamefont{{Bayliss}}},
  \bibinfo{author}{\bibfnamefont{B.~A.} \bibnamefont{{Benson}}},
  \bibinfo{author}{\bibfnamefont{L.~E.} \bibnamefont{{Bleem}}},
  \bibinfo{author}{\bibfnamefont{S.}~\bibnamefont{{Bocquet}}},
  \bibnamefont{et~al.}, \bibinfo{journal}{\mnras}
  \textbf{\bibinfo{volume}{448}}, \bibinfo{pages}{2085} (\bibinfo{year}{2015}),
  \eprint{1407.7520}.

\bibitem[{\citenamefont{{Chiu} et~al.}(2018)\citenamefont{{Chiu}, {Mohr},
  {McDonald}, {Bocquet}, {Desai}, {Klein}, {Israel}, {Ashby}, {Stanford},
  {Benson} et~al.}}]{Chiu18}
\bibinfo{author}{\bibfnamefont{I.}~\bibnamefont{{Chiu}}},
  \bibinfo{author}{\bibfnamefont{J.~J.} \bibnamefont{{Mohr}}},
  \bibinfo{author}{\bibfnamefont{M.}~\bibnamefont{{McDonald}}},
  \bibinfo{author}{\bibfnamefont{S.}~\bibnamefont{{Bocquet}}},
  \bibinfo{author}{\bibfnamefont{S.}~\bibnamefont{{Desai}}},
  \bibinfo{author}{\bibfnamefont{M.}~\bibnamefont{{Klein}}},
  \bibinfo{author}{\bibfnamefont{H.}~\bibnamefont{{Israel}}},
  \bibinfo{author}{\bibfnamefont{M.~L.~N.} \bibnamefont{{Ashby}}},
  \bibinfo{author}{\bibfnamefont{A.}~\bibnamefont{{Stanford}}},
  \bibinfo{author}{\bibfnamefont{B.~A.} \bibnamefont{{Benson}}},
  \bibnamefont{et~al.}, \bibinfo{journal}{\mnras}
  \textbf{\bibinfo{volume}{478}}, \bibinfo{pages}{3072} (\bibinfo{year}{2018}),
  \eprint{1711.00917}.

\bibitem[{\citenamefont{{De Martino} and
  {Atrio-Barandela}}(2016)}]{DeMartino16}
\bibinfo{author}{\bibfnamefont{I.}~\bibnamefont{{De Martino}}}
  \bibnamefont{and}
  \bibinfo{author}{\bibfnamefont{F.}~\bibnamefont{{Atrio-Barandela}}},
  \bibinfo{journal}{\mnras} \textbf{\bibinfo{volume}{461}},
  \bibinfo{pages}{3222} (\bibinfo{year}{2016}), \eprint{1606.04983}.

\bibitem[{\citenamefont{{Biffi} et~al.}(2014)\citenamefont{{Biffi},
  {Sembolini}, {De Petris}, {Valdarnini}, {Yepes}, and
  {Gottl{\"o}ber}}}]{Biffi14}
\bibinfo{author}{\bibfnamefont{V.}~\bibnamefont{{Biffi}}},
  \bibinfo{author}{\bibfnamefont{F.}~\bibnamefont{{Sembolini}}},
  \bibinfo{author}{\bibfnamefont{M.}~\bibnamefont{{De Petris}}},
  \bibinfo{author}{\bibfnamefont{R.}~\bibnamefont{{Valdarnini}}},
  \bibinfo{author}{\bibfnamefont{G.}~\bibnamefont{{Yepes}}}, \bibnamefont{and}
  \bibinfo{author}{\bibfnamefont{S.}~\bibnamefont{{Gottl{\"o}ber}}},
  \bibinfo{journal}{\mnras} \textbf{\bibinfo{volume}{439}},
  \bibinfo{pages}{588} (\bibinfo{year}{2014}), \eprint{1401.2992}.

\bibitem[{\citenamefont{{Bender} et~al.}(2016)\citenamefont{{Bender},
  {Kennedy}, {Ade}, {Basu}, {Bertoldi}, {Burkutean}, {Clarke}, {Dahlin},
  {Dobbs}, {Ferrusca} et~al.}}]{Bender}
\bibinfo{author}{\bibfnamefont{A.~N.} \bibnamefont{{Bender}}},
  \bibinfo{author}{\bibfnamefont{J.}~\bibnamefont{{Kennedy}}},
  \bibinfo{author}{\bibfnamefont{P.~A.~R.} \bibnamefont{{Ade}}},
  \bibinfo{author}{\bibfnamefont{K.}~\bibnamefont{{Basu}}},
  \bibinfo{author}{\bibfnamefont{F.}~\bibnamefont{{Bertoldi}}},
  \bibinfo{author}{\bibfnamefont{S.}~\bibnamefont{{Burkutean}}},
  \bibinfo{author}{\bibfnamefont{J.}~\bibnamefont{{Clarke}}},
  \bibinfo{author}{\bibfnamefont{D.}~\bibnamefont{{Dahlin}}},
  \bibinfo{author}{\bibfnamefont{M.}~\bibnamefont{{Dobbs}}},
  \bibinfo{author}{\bibfnamefont{D.}~\bibnamefont{{Ferrusca}}},
  \bibnamefont{et~al.}, \bibinfo{journal}{\mnras}
  \textbf{\bibinfo{volume}{460}}, \bibinfo{pages}{3432} (\bibinfo{year}{2016}),
  \eprint{1404.7103}.

\bibitem[{\citenamefont{{Zhu} et~al.}(2019)\citenamefont{{Zhu}, {Wang}, {Zhao},
  {Jia}, {Li}, and {Chen}}}]{Zhu}
\bibinfo{author}{\bibfnamefont{Y.}~\bibnamefont{{Zhu}}},
  \bibinfo{author}{\bibfnamefont{Y.-H.} \bibnamefont{{Wang}}},
  \bibinfo{author}{\bibfnamefont{H.-H.} \bibnamefont{{Zhao}}},
  \bibinfo{author}{\bibfnamefont{S.-M.} \bibnamefont{{Jia}}},
  \bibinfo{author}{\bibfnamefont{C.-K.} \bibnamefont{{Li}}}, \bibnamefont{and}
  \bibinfo{author}{\bibfnamefont{Y.}~\bibnamefont{{Chen}}},
  \bibinfo{journal}{Research in Astronomy and Astrophysics}
  \textbf{\bibinfo{volume}{19}}, \bibinfo{eid}{104} (\bibinfo{year}{2019}),
  \eprint{1902.07507}.

\bibitem[{\citenamefont{{Pratt} and {Bregman}}(2020)}]{Pratt}
\bibinfo{author}{\bibfnamefont{C.~T.} \bibnamefont{{Pratt}}} \bibnamefont{and}
  \bibinfo{author}{\bibfnamefont{J.~N.} \bibnamefont{{Bregman}}},
  \bibinfo{journal}{\apj} \textbf{\bibinfo{volume}{890}}, \bibinfo{eid}{156}
  (\bibinfo{year}{2020}), \eprint{2001.07802}.

\bibitem[{\citenamefont{{Henden} et~al.}(2018)\citenamefont{{Henden},
  {Puchwein}, {Shen}, and {Sijacki}}}]{Henden}
\bibinfo{author}{\bibfnamefont{N.~A.} \bibnamefont{{Henden}}},
  \bibinfo{author}{\bibfnamefont{E.}~\bibnamefont{{Puchwein}}},
  \bibinfo{author}{\bibfnamefont{S.}~\bibnamefont{{Shen}}}, \bibnamefont{and}
  \bibinfo{author}{\bibfnamefont{D.}~\bibnamefont{{Sijacki}}},
  \bibinfo{journal}{\mnras} \textbf{\bibinfo{volume}{479}},
  \bibinfo{pages}{5385} (\bibinfo{year}{2018}), \eprint{1804.05064}.

\bibitem[{\citenamefont{{Sunyaev} and {Zeldovich}}(1972)}]{sz}
\bibinfo{author}{\bibfnamefont{R.~A.} \bibnamefont{{Sunyaev}}}
  \bibnamefont{and} \bibinfo{author}{\bibfnamefont{Y.~B.}
  \bibnamefont{{Zeldovich}}}, \bibinfo{journal}{Comments on Astrophysics and
  Space Physics} \textbf{\bibinfo{volume}{4}}, \bibinfo{pages}{173}
  (\bibinfo{year}{1972}).

\bibitem[{\citenamefont{{Birkinshaw}}(1999)}]{birki}
\bibinfo{author}{\bibfnamefont{M.}~\bibnamefont{{Birkinshaw}}},
  \bibinfo{journal}{\physrep} \textbf{\bibinfo{volume}{310}},
  \bibinfo{pages}{97} (\bibinfo{year}{1999}), \eprint{astro-ph/9808050}.

\bibitem[{\citenamefont{{Carlstrom} et~al.}(2002)\citenamefont{{Carlstrom},
  {Holder}, and {Reese}}}]{carl}
\bibinfo{author}{\bibfnamefont{J.~E.} \bibnamefont{{Carlstrom}}},
  \bibinfo{author}{\bibfnamefont{G.~P.} \bibnamefont{{Holder}}},
  \bibnamefont{and} \bibinfo{author}{\bibfnamefont{E.~D.}
  \bibnamefont{{Reese}}}, \bibinfo{journal}{\araa}
  \textbf{\bibinfo{volume}{40}}, \bibinfo{pages}{643} (\bibinfo{year}{2002}),
  \eprint{astro-ph/0208192}.

\bibitem[{\citenamefont{{Mroczkowski} et~al.}(2019)\citenamefont{{Mroczkowski},
  {Nagai}, {Basu}, {Chluba}, {Sayers}, {Adam}, {Churazov}, {Crites}, {Di
  Mascolo}, {Eckert} et~al.}}]{SZ18}
\bibinfo{author}{\bibfnamefont{T.}~\bibnamefont{{Mroczkowski}}},
  \bibinfo{author}{\bibfnamefont{D.}~\bibnamefont{{Nagai}}},
  \bibinfo{author}{\bibfnamefont{K.}~\bibnamefont{{Basu}}},
  \bibinfo{author}{\bibfnamefont{J.}~\bibnamefont{{Chluba}}},
  \bibinfo{author}{\bibfnamefont{J.}~\bibnamefont{{Sayers}}},
  \bibinfo{author}{\bibfnamefont{R.}~\bibnamefont{{Adam}}},
  \bibinfo{author}{\bibfnamefont{E.}~\bibnamefont{{Churazov}}},
  \bibinfo{author}{\bibfnamefont{A.}~\bibnamefont{{Crites}}},
  \bibinfo{author}{\bibfnamefont{L.}~\bibnamefont{{Di Mascolo}}},
  \bibinfo{author}{\bibfnamefont{D.}~\bibnamefont{{Eckert}}},
  \bibnamefont{et~al.}, \bibinfo{journal}{\ssr} \textbf{\bibinfo{volume}{215}},
  \bibinfo{eid}{17} (\bibinfo{year}{2019}), \eprint{1811.02310}.

\bibitem[{\citenamefont{{Carlstrom} et~al.}(2011)\citenamefont{{Carlstrom},
  {Ade}, {Aird}, {Benson}, {Bleem}, {Busetti}, {Chang}, {Chauvin}, {Cho},
  {Crawford} et~al.}}]{Carlstrom}
\bibinfo{author}{\bibfnamefont{J.~E.} \bibnamefont{{Carlstrom}}},
  \bibinfo{author}{\bibfnamefont{P.~A.~R.} \bibnamefont{{Ade}}},
  \bibinfo{author}{\bibfnamefont{K.~A.} \bibnamefont{{Aird}}},
  \bibinfo{author}{\bibfnamefont{B.~A.} \bibnamefont{{Benson}}},
  \bibinfo{author}{\bibfnamefont{L.~E.} \bibnamefont{{Bleem}}},
  \bibinfo{author}{\bibfnamefont{S.}~\bibnamefont{{Busetti}}},
  \bibinfo{author}{\bibfnamefont{C.~L.} \bibnamefont{{Chang}}},
  \bibinfo{author}{\bibfnamefont{E.}~\bibnamefont{{Chauvin}}},
  \bibinfo{author}{\bibfnamefont{H.~M.} \bibnamefont{{Cho}}},
  \bibinfo{author}{\bibfnamefont{T.~M.} \bibnamefont{{Crawford}}},
  \bibnamefont{et~al.}, \bibinfo{journal}{\pasp}
  \textbf{\bibinfo{volume}{123}}, \bibinfo{pages}{568} (\bibinfo{year}{2011}),
  \eprint{0907.4445}.

\bibitem[{\citenamefont{{Swetz} et~al.}(2011)\citenamefont{{Swetz}, {Ade},
  {Amiri}, {Appel}, {Battistelli}, {Burger}, {Chervenak}, {Devlin}, {Dicker},
  {Doriese} et~al.}}]{Fowler}
\bibinfo{author}{\bibfnamefont{D.~S.} \bibnamefont{{Swetz}}},
  \bibinfo{author}{\bibfnamefont{P.~A.~R.} \bibnamefont{{Ade}}},
  \bibinfo{author}{\bibfnamefont{M.}~\bibnamefont{{Amiri}}},
  \bibinfo{author}{\bibfnamefont{J.~W.} \bibnamefont{{Appel}}},
  \bibinfo{author}{\bibfnamefont{E.~S.} \bibnamefont{{Battistelli}}},
  \bibinfo{author}{\bibfnamefont{B.}~\bibnamefont{{Burger}}},
  \bibinfo{author}{\bibfnamefont{J.}~\bibnamefont{{Chervenak}}},
  \bibinfo{author}{\bibfnamefont{M.~J.} \bibnamefont{{Devlin}}},
  \bibinfo{author}{\bibfnamefont{S.~R.} \bibnamefont{{Dicker}}},
  \bibinfo{author}{\bibfnamefont{W.~B.} \bibnamefont{{Doriese}}},
  \bibnamefont{et~al.}, \bibinfo{journal}{\apjs}
  \textbf{\bibinfo{volume}{194}}, \bibinfo{eid}{41} (\bibinfo{year}{2011}),
  \eprint{1007.0290}.

\bibitem[{\citenamefont{{Planck Collaboration}
  et~al.}(2016)\citenamefont{{Planck Collaboration}, {Ade}, {Aghanim},
  {Arnaud}, {Ashdown}, {Aumont}, {Baccigalupi}, {Banday}, {Barreiro}, {Barrena}
  et~al.}}]{Plancksz}
\bibinfo{author}{\bibnamefont{{Planck Collaboration}}},
  \bibinfo{author}{\bibfnamefont{P.~A.~R.} \bibnamefont{{Ade}}},
  \bibinfo{author}{\bibfnamefont{N.}~\bibnamefont{{Aghanim}}},
  \bibinfo{author}{\bibfnamefont{M.}~\bibnamefont{{Arnaud}}},
  \bibinfo{author}{\bibfnamefont{M.}~\bibnamefont{{Ashdown}}},
  \bibinfo{author}{\bibfnamefont{J.}~\bibnamefont{{Aumont}}},
  \bibinfo{author}{\bibfnamefont{C.}~\bibnamefont{{Baccigalupi}}},
  \bibinfo{author}{\bibfnamefont{A.~J.} \bibnamefont{{Banday}}},
  \bibinfo{author}{\bibfnamefont{R.~B.} \bibnamefont{{Barreiro}}},
  \bibinfo{author}{\bibfnamefont{R.}~\bibnamefont{{Barrena}}},
  \bibnamefont{et~al.}, \bibinfo{journal}{\aap} \textbf{\bibinfo{volume}{594}},
  \bibinfo{eid}{A27} (\bibinfo{year}{2016}), \eprint{1502.01598}.

\bibitem[{\citenamefont{{Kravtsov} et~al.}(2006)\citenamefont{{Kravtsov},
  {Vikhlinin}, and {Nagai}}}]{Kravtsov}
\bibinfo{author}{\bibfnamefont{A.~V.} \bibnamefont{{Kravtsov}}},
  \bibinfo{author}{\bibfnamefont{A.}~\bibnamefont{{Vikhlinin}}},
  \bibnamefont{and} \bibinfo{author}{\bibfnamefont{D.}~\bibnamefont{{Nagai}}},
  \bibinfo{journal}{\apj} \textbf{\bibinfo{volume}{650}}, \bibinfo{pages}{128}
  (\bibinfo{year}{2006}), \eprint{astro-ph/0603205}.

\bibitem[{\citenamefont{{Gon{\c{c}}alves}
  et~al.}(2012)\citenamefont{{Gon{\c{c}}alves}, {Holanda}, and
  {Alcaniz}}}]{Holanda11}
\bibinfo{author}{\bibfnamefont{R.~S.} \bibnamefont{{Gon{\c{c}}alves}}},
  \bibinfo{author}{\bibfnamefont{R.~F.~L.} \bibnamefont{{Holanda}}},
  \bibnamefont{and} \bibinfo{author}{\bibfnamefont{J.~S.}
  \bibnamefont{{Alcaniz}}}, \bibinfo{journal}{\mnras}
  \textbf{\bibinfo{volume}{420}}, \bibinfo{pages}{L43} (\bibinfo{year}{2012}),
  \eprint{1109.2790}.

\bibitem[{\citenamefont{{Hees} et~al.}(2014)\citenamefont{{Hees}, {Minazzoli},
  and {Larena}}}]{hees}
\bibinfo{author}{\bibfnamefont{A.}~\bibnamefont{{Hees}}},
  \bibinfo{author}{\bibfnamefont{O.}~\bibnamefont{{Minazzoli}}},
  \bibnamefont{and} \bibinfo{author}{\bibfnamefont{J.}~\bibnamefont{{Larena}}},
  \bibinfo{journal}{\prd} \textbf{\bibinfo{volume}{90}}, \bibinfo{eid}{124064}
  (\bibinfo{year}{2014}), \eprint{1406.6187}.

\bibitem[{\citenamefont{{Gon{\c{c}}alves}
  et~al.}(2020)\citenamefont{{Gon{\c{c}}alves}, {Landau}, {Alcaniz}, and
  {Holanda}}}]{rodrigo19}
\bibinfo{author}{\bibfnamefont{R.~S.} \bibnamefont{{Gon{\c{c}}alves}}},
  \bibinfo{author}{\bibfnamefont{S.}~\bibnamefont{{Landau}}},
  \bibinfo{author}{\bibfnamefont{J.~S.} \bibnamefont{{Alcaniz}}},
  \bibnamefont{and} \bibinfo{author}{\bibfnamefont{R.~F.~L.}
  \bibnamefont{{Holanda}}}, \bibinfo{journal}{\jcap}
  \textbf{\bibinfo{volume}{2020}}, \bibinfo{eid}{036} (\bibinfo{year}{2020}),
  \eprint{1907.02118}.

\bibitem[{\citenamefont{{Stanek} et~al.}(2010)\citenamefont{{Stanek}, {Rasia},
  {Evrard}, {Pearce}, and {Gazzola}}}]{stanek}
\bibinfo{author}{\bibfnamefont{R.}~\bibnamefont{{Stanek}}},
  \bibinfo{author}{\bibfnamefont{E.}~\bibnamefont{{Rasia}}},
  \bibinfo{author}{\bibfnamefont{A.~E.} \bibnamefont{{Evrard}}},
  \bibinfo{author}{\bibfnamefont{F.}~\bibnamefont{{Pearce}}}, \bibnamefont{and}
  \bibinfo{author}{\bibfnamefont{L.}~\bibnamefont{{Gazzola}}},
  \bibinfo{journal}{\apj} \textbf{\bibinfo{volume}{715}}, \bibinfo{pages}{1508}
  (\bibinfo{year}{2010}), \eprint{0910.1599}.

\bibitem[{\citenamefont{{Fabjan} et~al.}(2011)\citenamefont{{Fabjan},
  {Borgani}, {Rasia}, {Bonafede}, {Dolag}, {Murante}, and
  {Tornatore}}}]{fabjan}
\bibinfo{author}{\bibfnamefont{D.}~\bibnamefont{{Fabjan}}},
  \bibinfo{author}{\bibfnamefont{S.}~\bibnamefont{{Borgani}}},
  \bibinfo{author}{\bibfnamefont{E.}~\bibnamefont{{Rasia}}},
  \bibinfo{author}{\bibfnamefont{A.}~\bibnamefont{{Bonafede}}},
  \bibinfo{author}{\bibfnamefont{K.}~\bibnamefont{{Dolag}}},
  \bibinfo{author}{\bibfnamefont{G.}~\bibnamefont{{Murante}}},
  \bibnamefont{and}
  \bibinfo{author}{\bibfnamefont{L.}~\bibnamefont{{Tornatore}}},
  \bibinfo{journal}{\mnras} \textbf{\bibinfo{volume}{416}},
  \bibinfo{pages}{801} (\bibinfo{year}{2011}), \eprint{1102.2903}.

\bibitem[{\citenamefont{{Kay} et~al.}(2012)\citenamefont{{Kay}, {Peel},
  {Short}, {Thomas}, {Young}, {Battye}, {Liddle}, and {Pearce}}}]{kay}
\bibinfo{author}{\bibfnamefont{S.~T.} \bibnamefont{{Kay}}},
  \bibinfo{author}{\bibfnamefont{M.~W.} \bibnamefont{{Peel}}},
  \bibinfo{author}{\bibfnamefont{C.~J.} \bibnamefont{{Short}}},
  \bibinfo{author}{\bibfnamefont{P.~A.} \bibnamefont{{Thomas}}},
  \bibinfo{author}{\bibfnamefont{O.~E.} \bibnamefont{{Young}}},
  \bibinfo{author}{\bibfnamefont{R.~A.} \bibnamefont{{Battye}}},
  \bibinfo{author}{\bibfnamefont{A.~R.} \bibnamefont{{Liddle}}},
  \bibnamefont{and} \bibinfo{author}{\bibfnamefont{F.~R.}
  \bibnamefont{{Pearce}}}, \bibinfo{journal}{\mnras}
  \textbf{\bibinfo{volume}{422}}, \bibinfo{pages}{1999} (\bibinfo{year}{2012}),
  \eprint{1112.3769}.

\bibitem[{\citenamefont{{Loken} et~al.}(2002)\citenamefont{{Loken}, {Norman},
  {Nelson}, {Burns}, {Bryan}, and {Motl}}}]{Norman}
\bibinfo{author}{\bibfnamefont{C.}~\bibnamefont{{Loken}}},
  \bibinfo{author}{\bibfnamefont{M.~L.} \bibnamefont{{Norman}}},
  \bibinfo{author}{\bibfnamefont{E.}~\bibnamefont{{Nelson}}},
  \bibinfo{author}{\bibfnamefont{J.}~\bibnamefont{{Burns}}},
  \bibinfo{author}{\bibfnamefont{G.~L.} \bibnamefont{{Bryan}}},
  \bibnamefont{and} \bibinfo{author}{\bibfnamefont{P.}~\bibnamefont{{Motl}}},
  \bibinfo{journal}{\apj} \textbf{\bibinfo{volume}{579}}, \bibinfo{pages}{571}
  (\bibinfo{year}{2002}), \eprint{astro-ph/0207095}.

\bibitem[{\citenamefont{Martinelli et~al.}(2017)\citenamefont{Martinelli,
  Calabrese, and Martins}}]{erminia}
\bibinfo{author}{\bibfnamefont{M.}~\bibnamefont{Martinelli}},
  \bibinfo{author}{\bibfnamefont{E.}~\bibnamefont{Calabrese}},
  \bibnamefont{and} \bibinfo{author}{\bibfnamefont{C.~J.}
  \bibnamefont{Martins}}, in \emph{\bibinfo{booktitle}{{14th Marcel Grossmann
  Meeting on Recent Developments in Theoretical and Experimental General
  Relativity, Astrophysics, and Relativistic Field Theories}}}
  (\bibinfo{year}{2017}), vol.~\bibinfo{volume}{4}, pp.
  \bibinfo{pages}{3664--3669}.

\bibitem[{\citenamefont{{Planck Collaboration}
  et~al.}(2020)\citenamefont{{Planck Collaboration}, {Aghanim}, {Akrami},
  {Ashdown}, {Aumont}, {Baccigalupi}, {Ballardini}, {Banday}, {Barreiro},
  {Bartolo} et~al.}}]{planck18}
\bibinfo{author}{\bibnamefont{{Planck Collaboration}}},
  \bibinfo{author}{\bibfnamefont{N.}~\bibnamefont{{Aghanim}}},
  \bibinfo{author}{\bibfnamefont{Y.}~\bibnamefont{{Akrami}}},
  \bibinfo{author}{\bibfnamefont{M.}~\bibnamefont{{Ashdown}}},
  \bibinfo{author}{\bibfnamefont{J.}~\bibnamefont{{Aumont}}},
  \bibinfo{author}{\bibfnamefont{C.}~\bibnamefont{{Baccigalupi}}},
  \bibinfo{author}{\bibfnamefont{M.}~\bibnamefont{{Ballardini}}},
  \bibinfo{author}{\bibfnamefont{A.~J.} \bibnamefont{{Banday}}},
  \bibinfo{author}{\bibfnamefont{R.~B.} \bibnamefont{{Barreiro}}},
  \bibinfo{author}{\bibfnamefont{N.}~\bibnamefont{{Bartolo}}},
  \bibnamefont{et~al.}, \bibinfo{journal}{\aap} \textbf{\bibinfo{volume}{641}},
  \bibinfo{eid}{A6} (\bibinfo{year}{2020}), \eprint{1807.06209}.

\bibitem[{\citenamefont{{Astropy Collaboration}
  et~al.}(2018)\citenamefont{{Astropy Collaboration}, {Price-Whelan},
  {Sip{\H{o}}cz}, {G{\"u}nther}, {Lim}, {Crawford}, {Conseil}, {Shupe},
  {Craig}, {Dencheva} et~al.}}]{astropy}
\bibinfo{author}{\bibnamefont{{Astropy Collaboration}}},
  \bibinfo{author}{\bibfnamefont{A.~M.} \bibnamefont{{Price-Whelan}}},
  \bibinfo{author}{\bibfnamefont{B.~M.} \bibnamefont{{Sip{\H{o}}cz}}},
  \bibinfo{author}{\bibfnamefont{H.~M.} \bibnamefont{{G{\"u}nther}}},
  \bibinfo{author}{\bibfnamefont{P.~L.} \bibnamefont{{Lim}}},
  \bibinfo{author}{\bibfnamefont{S.~M.} \bibnamefont{{Crawford}}},
  \bibinfo{author}{\bibfnamefont{S.}~\bibnamefont{{Conseil}}},
  \bibinfo{author}{\bibfnamefont{D.~L.} \bibnamefont{{Shupe}}},
  \bibinfo{author}{\bibfnamefont{M.~W.} \bibnamefont{{Craig}}},
  \bibinfo{author}{\bibfnamefont{N.}~\bibnamefont{{Dencheva}}},
  \bibnamefont{et~al.}, \bibinfo{journal}{\aj} \textbf{\bibinfo{volume}{156}},
  \bibinfo{eid}{123} (\bibinfo{year}{2018}), \eprint{1801.02634}.

\bibitem[{\citenamefont{{Bulbul} et~al.}(2019)\citenamefont{{Bulbul}, {Chiu},
  {Mohr}, {McDonald}, {Benson}, {Bautz}, {Bayliss}, {Bleem}, {Brodwin},
  {Bocquet} et~al.}}]{Bulbul}
\bibinfo{author}{\bibfnamefont{E.}~\bibnamefont{{Bulbul}}},
  \bibinfo{author}{\bibfnamefont{I.~N.} \bibnamefont{{Chiu}}},
  \bibinfo{author}{\bibfnamefont{J.~J.} \bibnamefont{{Mohr}}},
  \bibinfo{author}{\bibfnamefont{M.}~\bibnamefont{{McDonald}}},
  \bibinfo{author}{\bibfnamefont{B.}~\bibnamefont{{Benson}}},
  \bibinfo{author}{\bibfnamefont{M.~W.} \bibnamefont{{Bautz}}},
  \bibinfo{author}{\bibfnamefont{M.}~\bibnamefont{{Bayliss}}},
  \bibinfo{author}{\bibfnamefont{L.}~\bibnamefont{{Bleem}}},
  \bibinfo{author}{\bibfnamefont{M.}~\bibnamefont{{Brodwin}}},
  \bibinfo{author}{\bibfnamefont{S.}~\bibnamefont{{Bocquet}}},
  \bibnamefont{et~al.}, \bibinfo{journal}{\apj} \textbf{\bibinfo{volume}{871}},
  \bibinfo{eid}{50} (\bibinfo{year}{2019}), \eprint{1807.02556}.

\bibitem[{\citenamefont{{Bleem} et~al.}(2015)\citenamefont{{Bleem}, {Stalder},
  {de Haan}, {Aird}, {Allen}, {Applegate}, {Ashby}, {Bautz}, {Bayliss},
  {Benson} et~al.}}]{Bleem15}
\bibinfo{author}{\bibfnamefont{L.~E.} \bibnamefont{{Bleem}}},
  \bibinfo{author}{\bibfnamefont{B.}~\bibnamefont{{Stalder}}},
  \bibinfo{author}{\bibfnamefont{T.}~\bibnamefont{{de Haan}}},
  \bibinfo{author}{\bibfnamefont{K.~A.} \bibnamefont{{Aird}}},
  \bibinfo{author}{\bibfnamefont{S.~W.} \bibnamefont{{Allen}}},
  \bibinfo{author}{\bibfnamefont{D.~E.} \bibnamefont{{Applegate}}},
  \bibinfo{author}{\bibfnamefont{M.~L.~N.} \bibnamefont{{Ashby}}},
  \bibinfo{author}{\bibfnamefont{M.}~\bibnamefont{{Bautz}}},
  \bibinfo{author}{\bibfnamefont{M.}~\bibnamefont{{Bayliss}}},
  \bibinfo{author}{\bibfnamefont{B.~A.} \bibnamefont{{Benson}}},
  \bibnamefont{et~al.}, \bibinfo{journal}{\apjs}
  \textbf{\bibinfo{volume}{216}}, \bibinfo{eid}{27} (\bibinfo{year}{2015}),
  \eprint{1409.0850}.

\bibitem[{\citenamefont{{Song} et~al.}(2012)\citenamefont{{Song}, {Zenteno},
  {Stalder}, {Desai}, {Bleem}, {Aird}, {Armstrong}, {Ashby}, {Bayliss}, {Bazin}
  et~al.}}]{Song}
\bibinfo{author}{\bibfnamefont{J.}~\bibnamefont{{Song}}},
  \bibinfo{author}{\bibfnamefont{A.}~\bibnamefont{{Zenteno}}},
  \bibinfo{author}{\bibfnamefont{B.}~\bibnamefont{{Stalder}}},
  \bibinfo{author}{\bibfnamefont{S.}~\bibnamefont{{Desai}}},
  \bibinfo{author}{\bibfnamefont{L.~E.} \bibnamefont{{Bleem}}},
  \bibinfo{author}{\bibfnamefont{K.~A.} \bibnamefont{{Aird}}},
  \bibinfo{author}{\bibfnamefont{R.}~\bibnamefont{{Armstrong}}},
  \bibinfo{author}{\bibfnamefont{M.~L.~N.} \bibnamefont{{Ashby}}},
  \bibinfo{author}{\bibfnamefont{M.}~\bibnamefont{{Bayliss}}},
  \bibinfo{author}{\bibfnamefont{G.}~\bibnamefont{{Bazin}}},
  \bibnamefont{et~al.}, \bibinfo{journal}{\apj} \textbf{\bibinfo{volume}{761}},
  \bibinfo{eid}{22} (\bibinfo{year}{2012}), \eprint{1207.4369}.

\bibitem[{\citenamefont{{Ruel} et~al.}(2014)\citenamefont{{Ruel}, {Bazin},
  {Bayliss}, {Brodwin}, {Foley}, {Stalder}, {Aird}, {Armstrong}, {Ashby},
  {Bautz} et~al.}}]{Ruel}
\bibinfo{author}{\bibfnamefont{J.}~\bibnamefont{{Ruel}}},
  \bibinfo{author}{\bibfnamefont{G.}~\bibnamefont{{Bazin}}},
  \bibinfo{author}{\bibfnamefont{M.}~\bibnamefont{{Bayliss}}},
  \bibinfo{author}{\bibfnamefont{M.}~\bibnamefont{{Brodwin}}},
  \bibinfo{author}{\bibfnamefont{R.~J.} \bibnamefont{{Foley}}},
  \bibinfo{author}{\bibfnamefont{B.}~\bibnamefont{{Stalder}}},
  \bibinfo{author}{\bibfnamefont{K.~A.} \bibnamefont{{Aird}}},
  \bibinfo{author}{\bibfnamefont{R.}~\bibnamefont{{Armstrong}}},
  \bibinfo{author}{\bibfnamefont{M.~L.~N.} \bibnamefont{{Ashby}}},
  \bibinfo{author}{\bibfnamefont{M.}~\bibnamefont{{Bautz}}},
  \bibnamefont{et~al.}, \bibinfo{journal}{\apj} \textbf{\bibinfo{volume}{792}},
  \bibinfo{eid}{45} (\bibinfo{year}{2014}), \eprint{1311.4953}.

\bibitem[{\citenamefont{{Bayliss} et~al.}(2017)\citenamefont{{Bayliss},
  {Zengo}, {Ruel}, {Benson}, {Bleem}, {Bocquet}, {Bulbul}, {Brodwin},
  {Capasso}, {Chiu} et~al.}}]{Bayliss}
\bibinfo{author}{\bibfnamefont{M.~B.} \bibnamefont{{Bayliss}}},
  \bibinfo{author}{\bibfnamefont{K.}~\bibnamefont{{Zengo}}},
  \bibinfo{author}{\bibfnamefont{J.}~\bibnamefont{{Ruel}}},
  \bibinfo{author}{\bibfnamefont{B.~A.} \bibnamefont{{Benson}}},
  \bibinfo{author}{\bibfnamefont{L.~E.} \bibnamefont{{Bleem}}},
  \bibinfo{author}{\bibfnamefont{S.}~\bibnamefont{{Bocquet}}},
  \bibinfo{author}{\bibfnamefont{E.}~\bibnamefont{{Bulbul}}},
  \bibinfo{author}{\bibfnamefont{M.}~\bibnamefont{{Brodwin}}},
  \bibinfo{author}{\bibfnamefont{R.}~\bibnamefont{{Capasso}}},
  \bibinfo{author}{\bibfnamefont{I.~n.} \bibnamefont{{Chiu}}},
  \bibnamefont{et~al.}, \bibinfo{journal}{\apj} \textbf{\bibinfo{volume}{837}},
  \bibinfo{eid}{88} (\bibinfo{year}{2017}), \eprint{1612.02827}.

\bibitem[{\citenamefont{{Desai} et~al.}(2012)\citenamefont{{Desai},
  {Armstrong}, {Mohr}, {Semler}, {Liu}, {Bertin}, {Allam}, {Barkhouse},
  {Bazin}, {Buckley-Geer} et~al.}}]{Desai12}
\bibinfo{author}{\bibfnamefont{S.}~\bibnamefont{{Desai}}},
  \bibinfo{author}{\bibfnamefont{R.}~\bibnamefont{{Armstrong}}},
  \bibinfo{author}{\bibfnamefont{J.~J.} \bibnamefont{{Mohr}}},
  \bibinfo{author}{\bibfnamefont{D.~R.} \bibnamefont{{Semler}}},
  \bibinfo{author}{\bibfnamefont{J.}~\bibnamefont{{Liu}}},
  \bibinfo{author}{\bibfnamefont{E.}~\bibnamefont{{Bertin}}},
  \bibinfo{author}{\bibfnamefont{S.~S.} \bibnamefont{{Allam}}},
  \bibinfo{author}{\bibfnamefont{W.~A.} \bibnamefont{{Barkhouse}}},
  \bibinfo{author}{\bibfnamefont{G.}~\bibnamefont{{Bazin}}},
  \bibinfo{author}{\bibfnamefont{E.~J.} \bibnamefont{{Buckley-Geer}}},
  \bibnamefont{et~al.}, \bibinfo{journal}{\apj} \textbf{\bibinfo{volume}{757}},
  \bibinfo{eid}{83} (\bibinfo{year}{2012}), \eprint{1204.1210}.

\bibitem[{\citenamefont{{Saro} et~al.}(2015)\citenamefont{{Saro}, {Bocquet},
  {Rozo}, {Benson}, {Mohr}, {Rykoff}, {Soares-Santos}, {Bleem}, {Dodelson},
  {Melchior} et~al.}}]{Saro}
\bibinfo{author}{\bibfnamefont{A.}~\bibnamefont{{Saro}}},
  \bibinfo{author}{\bibfnamefont{S.}~\bibnamefont{{Bocquet}}},
  \bibinfo{author}{\bibfnamefont{E.}~\bibnamefont{{Rozo}}},
  \bibinfo{author}{\bibfnamefont{B.~A.} \bibnamefont{{Benson}}},
  \bibinfo{author}{\bibfnamefont{J.}~\bibnamefont{{Mohr}}},
  \bibinfo{author}{\bibfnamefont{E.~S.} \bibnamefont{{Rykoff}}},
  \bibinfo{author}{\bibfnamefont{M.}~\bibnamefont{{Soares-Santos}}},
  \bibinfo{author}{\bibfnamefont{L.}~\bibnamefont{{Bleem}}},
  \bibinfo{author}{\bibfnamefont{S.}~\bibnamefont{{Dodelson}}},
  \bibinfo{author}{\bibfnamefont{P.}~\bibnamefont{{Melchior}}},
  \bibnamefont{et~al.}, \bibinfo{journal}{\mnras}
  \textbf{\bibinfo{volume}{454}}, \bibinfo{pages}{2305} (\bibinfo{year}{2015}),
  \eprint{1506.07814}.

\bibitem[{\citenamefont{{Bocquet} et~al.}(2019)\citenamefont{{Bocquet},
  {Dietrich}, {Schrabback}, {Bleem}, {Klein}, {Allen}, {Applegate}, {Ashby},
  {Bautz}, {Bayliss} et~al.}}]{Bocquet19}
\bibinfo{author}{\bibfnamefont{S.}~\bibnamefont{{Bocquet}}},
  \bibinfo{author}{\bibfnamefont{J.~P.} \bibnamefont{{Dietrich}}},
  \bibinfo{author}{\bibfnamefont{T.}~\bibnamefont{{Schrabback}}},
  \bibinfo{author}{\bibfnamefont{L.~E.} \bibnamefont{{Bleem}}},
  \bibinfo{author}{\bibfnamefont{M.}~\bibnamefont{{Klein}}},
  \bibinfo{author}{\bibfnamefont{S.~W.} \bibnamefont{{Allen}}},
  \bibinfo{author}{\bibfnamefont{D.~E.} \bibnamefont{{Applegate}}},
  \bibinfo{author}{\bibfnamefont{M.~L.~N.} \bibnamefont{{Ashby}}},
  \bibinfo{author}{\bibfnamefont{M.}~\bibnamefont{{Bautz}}},
  \bibinfo{author}{\bibfnamefont{M.}~\bibnamefont{{Bayliss}}},
  \bibnamefont{et~al.}, \bibinfo{journal}{\apj} \textbf{\bibinfo{volume}{878}},
  \bibinfo{eid}{55} (\bibinfo{year}{2019}), \eprint{1812.01679}.

\bibitem[{\citenamefont{{Arnaud} et~al.}(2010)\citenamefont{{Arnaud}, {Pratt},
  {Piffaretti}, {B{\"o}hringer}, {Croston}, and {Pointecouteau}}}]{Arnaud}
\bibinfo{author}{\bibfnamefont{M.}~\bibnamefont{{Arnaud}}},
  \bibinfo{author}{\bibfnamefont{G.~W.} \bibnamefont{{Pratt}}},
  \bibinfo{author}{\bibfnamefont{R.}~\bibnamefont{{Piffaretti}}},
  \bibinfo{author}{\bibfnamefont{H.}~\bibnamefont{{B{\"o}hringer}}},
  \bibinfo{author}{\bibfnamefont{J.~H.} \bibnamefont{{Croston}}},
  \bibnamefont{and}
  \bibinfo{author}{\bibfnamefont{E.}~\bibnamefont{{Pointecouteau}}},
  \bibinfo{journal}{\aap} \textbf{\bibinfo{volume}{517}}, \bibinfo{eid}{A92}
  (\bibinfo{year}{2010}), \eprint{0910.1234}.

\bibitem[{\citenamefont{{Melin} et~al.}(2011)\citenamefont{{Melin}, {Bartlett},
  {Delabrouille}, {Arnaud}, {Piffaretti}, and {Pratt}}}]{Melin}
\bibinfo{author}{\bibfnamefont{J.~B.} \bibnamefont{{Melin}}},
  \bibinfo{author}{\bibfnamefont{J.~G.} \bibnamefont{{Bartlett}}},
  \bibinfo{author}{\bibfnamefont{J.}~\bibnamefont{{Delabrouille}}},
  \bibinfo{author}{\bibfnamefont{M.}~\bibnamefont{{Arnaud}}},
  \bibinfo{author}{\bibfnamefont{R.}~\bibnamefont{{Piffaretti}}},
  \bibnamefont{and} \bibinfo{author}{\bibfnamefont{G.~W.}
  \bibnamefont{{Pratt}}}, \bibinfo{journal}{\aap}
  \textbf{\bibinfo{volume}{525}}, \bibinfo{eid}{A139} (\bibinfo{year}{2011}),
  \eprint{1001.0871}.

\bibitem[{\citenamefont{{Saro} et~al.}(2014)\citenamefont{{Saro}, {Liu},
  {Mohr}, {Aird}, {Ashby}, {Bayliss}, {Benson}, {Bleem}, {Bocquet}, {Brodwin}
  et~al.}}]{Saro13}
\bibinfo{author}{\bibfnamefont{A.}~\bibnamefont{{Saro}}},
  \bibinfo{author}{\bibfnamefont{J.}~\bibnamefont{{Liu}}},
  \bibinfo{author}{\bibfnamefont{J.~J.} \bibnamefont{{Mohr}}},
  \bibinfo{author}{\bibfnamefont{K.~A.} \bibnamefont{{Aird}}},
  \bibinfo{author}{\bibfnamefont{M.~L.~N.} \bibnamefont{{Ashby}}},
  \bibinfo{author}{\bibfnamefont{M.}~\bibnamefont{{Bayliss}}},
  \bibinfo{author}{\bibfnamefont{B.~A.} \bibnamefont{{Benson}}},
  \bibinfo{author}{\bibfnamefont{L.~E.} \bibnamefont{{Bleem}}},
  \bibinfo{author}{\bibfnamefont{S.}~\bibnamefont{{Bocquet}}},
  \bibinfo{author}{\bibfnamefont{M.}~\bibnamefont{{Brodwin}}},
  \bibnamefont{et~al.}, \bibinfo{journal}{\mnras}
  \textbf{\bibinfo{volume}{440}}, \bibinfo{pages}{2610} (\bibinfo{year}{2014}),
  \eprint{1312.2462}.

\bibitem[{\citenamefont{{McDonald} et~al.}(2013)\citenamefont{{McDonald},
  {Benson}, {Vikhlinin}, {Stalder}, {Bleem}, {de Haan}, {Lin}, {Aird}, {Ashby},
  {Bautz} et~al.}}]{McDonald13}
\bibinfo{author}{\bibfnamefont{M.}~\bibnamefont{{McDonald}}},
  \bibinfo{author}{\bibfnamefont{B.~A.} \bibnamefont{{Benson}}},
  \bibinfo{author}{\bibfnamefont{A.}~\bibnamefont{{Vikhlinin}}},
  \bibinfo{author}{\bibfnamefont{B.}~\bibnamefont{{Stalder}}},
  \bibinfo{author}{\bibfnamefont{L.~E.} \bibnamefont{{Bleem}}},
  \bibinfo{author}{\bibfnamefont{T.}~\bibnamefont{{de Haan}}},
  \bibinfo{author}{\bibfnamefont{H.~W.} \bibnamefont{{Lin}}},
  \bibinfo{author}{\bibfnamefont{K.~A.} \bibnamefont{{Aird}}},
  \bibinfo{author}{\bibfnamefont{M.~L.~N.} \bibnamefont{{Ashby}}},
  \bibinfo{author}{\bibfnamefont{M.~W.} \bibnamefont{{Bautz}}},
  \bibnamefont{et~al.}, \bibinfo{journal}{\apj} \textbf{\bibinfo{volume}{774}},
  \bibinfo{eid}{23} (\bibinfo{year}{2013}), \eprint{1305.2915}.

\bibitem[{\citenamefont{{McDonald} et~al.}(2014)\citenamefont{{McDonald},
  {Benson}, {Vikhlinin}, {Aird}, {Allen}, {Bautz}, {Bayliss}, {Bleem},
  {Bocquet}, {Brodwin} et~al.}}]{McDonald14}
\bibinfo{author}{\bibfnamefont{M.}~\bibnamefont{{McDonald}}},
  \bibinfo{author}{\bibfnamefont{B.~A.} \bibnamefont{{Benson}}},
  \bibinfo{author}{\bibfnamefont{A.}~\bibnamefont{{Vikhlinin}}},
  \bibinfo{author}{\bibfnamefont{K.~A.} \bibnamefont{{Aird}}},
  \bibinfo{author}{\bibfnamefont{S.~W.} \bibnamefont{{Allen}}},
  \bibinfo{author}{\bibfnamefont{M.}~\bibnamefont{{Bautz}}},
  \bibinfo{author}{\bibfnamefont{M.}~\bibnamefont{{Bayliss}}},
  \bibinfo{author}{\bibfnamefont{L.~E.} \bibnamefont{{Bleem}}},
  \bibinfo{author}{\bibfnamefont{S.}~\bibnamefont{{Bocquet}}},
  \bibinfo{author}{\bibfnamefont{M.}~\bibnamefont{{Brodwin}}},
  \bibnamefont{et~al.}, \bibinfo{journal}{\apj} \textbf{\bibinfo{volume}{794}},
  \bibinfo{eid}{67} (\bibinfo{year}{2014}), \eprint{1404.6250}.

\bibitem[{\citenamefont{{Semler} et~al.}(2012)\citenamefont{{Semler},
  {{\v{S}}uhada}, {Aird}, {Ashby}, {Bautz}, {Bayliss}, {Bazin}, {Bocquet},
  {Benson}, {Bleem} et~al.}}]{Semler}
\bibinfo{author}{\bibfnamefont{D.~R.} \bibnamefont{{Semler}}},
  \bibinfo{author}{\bibfnamefont{R.}~\bibnamefont{{{\v{S}}uhada}}},
  \bibinfo{author}{\bibfnamefont{K.~A.} \bibnamefont{{Aird}}},
  \bibinfo{author}{\bibfnamefont{M.~L.~N.} \bibnamefont{{Ashby}}},
  \bibinfo{author}{\bibfnamefont{M.}~\bibnamefont{{Bautz}}},
  \bibinfo{author}{\bibfnamefont{M.}~\bibnamefont{{Bayliss}}},
  \bibinfo{author}{\bibfnamefont{G.}~\bibnamefont{{Bazin}}},
  \bibinfo{author}{\bibfnamefont{S.}~\bibnamefont{{Bocquet}}},
  \bibinfo{author}{\bibfnamefont{B.~A.} \bibnamefont{{Benson}}},
  \bibinfo{author}{\bibfnamefont{L.~E.} \bibnamefont{{Bleem}}},
  \bibnamefont{et~al.}, \bibinfo{journal}{\apj} \textbf{\bibinfo{volume}{761}},
  \bibinfo{eid}{183} (\bibinfo{year}{2012}), \eprint{1208.3368}.

\bibitem[{\citenamefont{{McDonald} et~al.}(2017)\citenamefont{{McDonald},
  {Allen}, {Bayliss}, {Benson}, {Bleem}, {Brodwin}, {Bulbul}, {Carlstrom},
  {Forman}, {Hlavacek-Larrondo} et~al.}}]{McDonald17}
\bibinfo{author}{\bibfnamefont{M.}~\bibnamefont{{McDonald}}},
  \bibinfo{author}{\bibfnamefont{S.~W.} \bibnamefont{{Allen}}},
  \bibinfo{author}{\bibfnamefont{M.}~\bibnamefont{{Bayliss}}},
  \bibinfo{author}{\bibfnamefont{B.~A.} \bibnamefont{{Benson}}},
  \bibinfo{author}{\bibfnamefont{L.~E.} \bibnamefont{{Bleem}}},
  \bibinfo{author}{\bibfnamefont{M.}~\bibnamefont{{Brodwin}}},
  \bibinfo{author}{\bibfnamefont{E.}~\bibnamefont{{Bulbul}}},
  \bibinfo{author}{\bibfnamefont{J.~E.} \bibnamefont{{Carlstrom}}},
  \bibinfo{author}{\bibfnamefont{W.~R.} \bibnamefont{{Forman}}},
  \bibinfo{author}{\bibfnamefont{J.}~\bibnamefont{{Hlavacek-Larrondo}}},
  \bibnamefont{et~al.}, \bibinfo{journal}{\apj} \textbf{\bibinfo{volume}{843}},
  \bibinfo{eid}{28} (\bibinfo{year}{2017}), \eprint{1702.05094}.

\bibitem[{\citenamefont{{McDonald} et~al.}(2019)\citenamefont{{McDonald},
  {Allen}, {Hlavacek-Larrondo}, {Mantz}, {Bayliss}, {Benson}, {Brodwin},
  {Bulbul}, {Canning}, {Chiu} et~al.}}]{McDonald19}
\bibinfo{author}{\bibfnamefont{M.}~\bibnamefont{{McDonald}}},
  \bibinfo{author}{\bibfnamefont{S.~W.} \bibnamefont{{Allen}}},
  \bibinfo{author}{\bibfnamefont{J.}~\bibnamefont{{Hlavacek-Larrondo}}},
  \bibinfo{author}{\bibfnamefont{A.~B.} \bibnamefont{{Mantz}}},
  \bibinfo{author}{\bibfnamefont{M.}~\bibnamefont{{Bayliss}}},
  \bibinfo{author}{\bibfnamefont{B.~A.} \bibnamefont{{Benson}}},
  \bibinfo{author}{\bibfnamefont{M.}~\bibnamefont{{Brodwin}}},
  \bibinfo{author}{\bibfnamefont{E.}~\bibnamefont{{Bulbul}}},
  \bibinfo{author}{\bibfnamefont{R.~E.~A.} \bibnamefont{{Canning}}},
  \bibinfo{author}{\bibfnamefont{I.}~\bibnamefont{{Chiu}}},
  \bibnamefont{et~al.}, \bibinfo{journal}{\apj} \textbf{\bibinfo{volume}{870}},
  \bibinfo{eid}{85} (\bibinfo{year}{2019}), \eprint{1809.09104}.

\bibitem[{\citenamefont{{Kelly}}(2007)}]{Kelly}
\bibinfo{author}{\bibfnamefont{B.~C.} \bibnamefont{{Kelly}}},
  \bibinfo{journal}{\apj} \textbf{\bibinfo{volume}{665}}, \bibinfo{pages}{1489}
  (\bibinfo{year}{2007}), \eprint{0705.2774}.

\bibitem[{\citenamefont{{Hogg} et~al.}(2010)\citenamefont{{Hogg}, {Bovy}, and
  {Lang}}}]{Hogg10}
\bibinfo{author}{\bibfnamefont{D.~W.} \bibnamefont{{Hogg}}},
  \bibinfo{author}{\bibfnamefont{J.}~\bibnamefont{{Bovy}}}, \bibnamefont{and}
  \bibinfo{author}{\bibfnamefont{D.}~\bibnamefont{{Lang}}},
  \bibinfo{journal}{arXiv e-prints} \bibinfo{eid}{arXiv:1008.4686}
  (\bibinfo{year}{2010}), \eprint{1008.4686}.

\bibitem[{\citenamefont{{Chiu} et~al.}(2020)\citenamefont{{Chiu}, {Umetsu},
  {Murata}, {Medezinski}, and {Oguri}}}]{Chiu20}
\bibinfo{author}{\bibfnamefont{I.~N.} \bibnamefont{{Chiu}}},
  \bibinfo{author}{\bibfnamefont{K.}~\bibnamefont{{Umetsu}}},
  \bibinfo{author}{\bibfnamefont{R.}~\bibnamefont{{Murata}}},
  \bibinfo{author}{\bibfnamefont{E.}~\bibnamefont{{Medezinski}}},
  \bibnamefont{and} \bibinfo{author}{\bibfnamefont{M.}~\bibnamefont{{Oguri}}},
  \bibinfo{journal}{\mnras} \textbf{\bibinfo{volume}{495}},
  \bibinfo{pages}{428} (\bibinfo{year}{2020}), \eprint{1909.02042}.

\bibitem[{\citenamefont{{Tian} et~al.}(2020)\citenamefont{{Tian}, {Umetsu},
  {Ko}, {Donahue}, and {Chiu}}}]{Tian}
\bibinfo{author}{\bibfnamefont{Y.}~\bibnamefont{{Tian}}},
  \bibinfo{author}{\bibfnamefont{K.}~\bibnamefont{{Umetsu}}},
  \bibinfo{author}{\bibfnamefont{C.-M.} \bibnamefont{{Ko}}},
  \bibinfo{author}{\bibfnamefont{M.}~\bibnamefont{{Donahue}}},
  \bibnamefont{and} \bibinfo{author}{\bibfnamefont{I.~N.}
  \bibnamefont{{Chiu}}}, \bibinfo{journal}{\apj}
  \textbf{\bibinfo{volume}{896}}, \bibinfo{eid}{70} (\bibinfo{year}{2020}),
  \eprint{2001.08340}.

\bibitem[{\citenamefont{Foreman-Mackey
  et~al.}(2013)\citenamefont{Foreman-Mackey, Hogg, Lang, and Goodman}}]{emcee}
\bibinfo{author}{\bibfnamefont{D.}~\bibnamefont{Foreman-Mackey}},
  \bibinfo{author}{\bibfnamefont{D.~W.} \bibnamefont{Hogg}},
  \bibinfo{author}{\bibfnamefont{D.}~\bibnamefont{Lang}}, \bibnamefont{and}
  \bibinfo{author}{\bibfnamefont{J.}~\bibnamefont{Goodman}},
  \bibinfo{journal}{Publ. Astron. Soc. Pac.} \textbf{\bibinfo{volume}{125}},
  \bibinfo{pages}{306} (\bibinfo{year}{2013}), \eprint{1202.3665}.

\bibitem[{\citenamefont{{Martins} and {Vacher}}(2019)}]{Vacher}
\bibinfo{author}{\bibfnamefont{C.~J.~A.~P.} \bibnamefont{{Martins}}}
  \bibnamefont{and} \bibinfo{author}{\bibfnamefont{L.}~\bibnamefont{{Vacher}}},
  \bibinfo{journal}{\prd} \textbf{\bibinfo{volume}{100}}, \bibinfo{eid}{123514}
  (\bibinfo{year}{2019}), \eprint{1911.10821}.

\bibitem[{\citenamefont{Benson et~al.}(2014)}]{Benson14}
\bibinfo{author}{\bibfnamefont{B.}~\bibnamefont{Benson}} \bibnamefont{et~al.}
  (\bibinfo{collaboration}{SPT-3G}), \bibinfo{journal}{Proc. SPIE Int. Soc.
  Opt. Eng.} \textbf{\bibinfo{volume}{9153}}, \bibinfo{pages}{91531P}
  (\bibinfo{year}{2014}), \eprint{1407.2973}.

\bibitem[{\citenamefont{{Bleem} et~al.}(2020)\citenamefont{{Bleem}, {Bocquet},
  {Stalder}, {Gladders}, {Ade}, {Allen}, {Anderson}, {Annis}, {Ashby},
  {Austermann} et~al.}}]{Bleem19}
\bibinfo{author}{\bibfnamefont{L.~E.} \bibnamefont{{Bleem}}},
  \bibinfo{author}{\bibfnamefont{S.}~\bibnamefont{{Bocquet}}},
  \bibinfo{author}{\bibfnamefont{B.}~\bibnamefont{{Stalder}}},
  \bibinfo{author}{\bibfnamefont{M.~D.} \bibnamefont{{Gladders}}},
  \bibinfo{author}{\bibfnamefont{P.~A.~R.} \bibnamefont{{Ade}}},
  \bibinfo{author}{\bibfnamefont{S.~W.} \bibnamefont{{Allen}}},
  \bibinfo{author}{\bibfnamefont{A.~J.} \bibnamefont{{Anderson}}},
  \bibinfo{author}{\bibfnamefont{J.}~\bibnamefont{{Annis}}},
  \bibinfo{author}{\bibfnamefont{M.~L.~N.} \bibnamefont{{Ashby}}},
  \bibinfo{author}{\bibfnamefont{J.~E.} \bibnamefont{{Austermann}}},
  \bibnamefont{et~al.}, \bibinfo{journal}{\apjs}
  \textbf{\bibinfo{volume}{247}}, \bibinfo{eid}{25} (\bibinfo{year}{2020}),
  \eprint{1910.04121}.

\bibitem[{\citenamefont{{Thornton} et~al.}(2016)\citenamefont{{Thornton},
  {Ade}, {Aiola}, {Angil{\`e}}, {Amiri}, {Beall}, {Becker}, {Cho}, {Choi},
  {Corlies} et~al.}}]{Thornton}
\bibinfo{author}{\bibfnamefont{R.~J.} \bibnamefont{{Thornton}}},
  \bibinfo{author}{\bibfnamefont{P.~A.~R.} \bibnamefont{{Ade}}},
  \bibinfo{author}{\bibfnamefont{S.}~\bibnamefont{{Aiola}}},
  \bibinfo{author}{\bibfnamefont{F.~E.} \bibnamefont{{Angil{\`e}}}},
  \bibinfo{author}{\bibfnamefont{M.}~\bibnamefont{{Amiri}}},
  \bibinfo{author}{\bibfnamefont{J.~A.} \bibnamefont{{Beall}}},
  \bibinfo{author}{\bibfnamefont{D.~T.} \bibnamefont{{Becker}}},
  \bibinfo{author}{\bibfnamefont{H.~M.} \bibnamefont{{Cho}}},
  \bibinfo{author}{\bibfnamefont{S.~K.} \bibnamefont{{Choi}}},
  \bibinfo{author}{\bibfnamefont{P.}~\bibnamefont{{Corlies}}},
  \bibnamefont{et~al.}, \bibinfo{journal}{\apjs}
  \textbf{\bibinfo{volume}{227}}, \bibinfo{eid}{21} (\bibinfo{year}{2016}),
  \eprint{1605.06569}.

\bibitem[{\citenamefont{{Hofmann} et~al.}(2017)\citenamefont{{Hofmann},
  {Sanders}, {Clerc}, {Nandra}, {Ridl}, {Dennerl}, {Ramos-Ceja}, {Finoguenov},
  and {Reiprich}}}]{erosita}
\bibinfo{author}{\bibfnamefont{F.}~\bibnamefont{{Hofmann}}},
  \bibinfo{author}{\bibfnamefont{J.~S.} \bibnamefont{{Sanders}}},
  \bibinfo{author}{\bibfnamefont{N.}~\bibnamefont{{Clerc}}},
  \bibinfo{author}{\bibfnamefont{K.}~\bibnamefont{{Nandra}}},
  \bibinfo{author}{\bibfnamefont{J.}~\bibnamefont{{Ridl}}},
  \bibinfo{author}{\bibfnamefont{K.}~\bibnamefont{{Dennerl}}},
  \bibinfo{author}{\bibfnamefont{M.}~\bibnamefont{{Ramos-Ceja}}},
  \bibinfo{author}{\bibfnamefont{A.}~\bibnamefont{{Finoguenov}}},
  \bibnamefont{and} \bibinfo{author}{\bibfnamefont{T.~H.}
  \bibnamefont{{Reiprich}}}, \bibinfo{journal}{\aap}
  \textbf{\bibinfo{volume}{606}}, \bibinfo{eid}{A118} (\bibinfo{year}{2017}),
  \eprint{1708.05205}.

\end{thebibliography}

\end{document}